\documentclass[conference]{IEEEtran}
\ifCLASSINFOpdf
\else
\fi
\usepackage[latin1]{inputenc}
\usepackage{amsmath}
\usepackage{amssymb}
\usepackage{amsfonts}
\usepackage{hhline}
\usepackage{algorithm}
\usepackage{algorithmic}
\usepackage{graphicx}
\usepackage{url}
\newtheorem{definition}{Definition}
\newtheorem{theorem}{Theorem}

\begin{document}
%
\title{Universal Approximation \\of Edge Density in Large Graphs}

\author{\IEEEauthorblockN{Marc Boullé}
\IEEEauthorblockA{Orange Labs\\
22300 Lannion, France\\
Email: marc.boulle@orange.com}}


%


\maketitle

\begin{abstract}
In this paper, we present a novel way to summarize the structure of large graphs, based on non-parametric estimation of edge density in directed multigraphs. 
Following coclustering approach, we use a clustering of the vertices, with a piecewise constant estimation of the density of the edges across the clusters, and address the problem of automatically and reliably inferring the number of clusters, which is the granularity of the coclustering.
We use a model selection technique with data-dependent prior and obtain an exact evaluation criterion for the posterior probability of edge density estimation models.
We demonstrate, both theoretically and empirically, that our data-dependent modeling technique is consistent, resilient to noise, valid non asymptotically and asymptotically behaves as an universal approximator of the true edge density in directed multigraphs.
We evaluate our method using artificial graphs and present its practical interest on real world graphs.
The method is both robust and scalable. It is able to extract insightful patterns in the unsupervised learning setting and to provide state of the art accuracy when used as a preparation step for supervised learning.
\end{abstract}


%
\IEEEpeerreviewmaketitle

\section{Introduction}
\label{secIntroduction}

With the recent availability of much network data, such as world wide web, social networks, phone call networks, science collaboration graphs \cite{AlbertEtAl02,AggarwalEtAl10}, there is a renewed interest for the graph partitioning problem, especially for the automatic discovery of community structures in large networks \cite{BlondelEtAl08,LeskovecEtAl09,JinEtAl11}.
Beyond clustering approaches, coclustering approaches aim at summarizing the relation between two entities in a many-to-many relationship.
Such a relation can be represented as a graph, where the source and target vertices represent entities and the edges stand for relations between entities.
A coclustering model provides a summary of a graph by grouping source vertices and target vertices.
For example, in \emph{market analysis}, the source vertices of the graph represent customers, the target vertices represent products and there is one edge each time a customer has purchased a product.
A coclustering model summarizes the dataset by grouping customers that have purchased approximately the same products and grouping products that have been purchased by approximately the same customers.
Coclustering models have been applied to many other domains, such as \emph{information retrieval} (the entities are documents and their words in a text corpus), \emph{web log analysis} (cookies and their visited web pages), \emph{web structure analysis} (web pages with  hyper-links between them) or \emph{telecommunication network} (the call detail records stand for the edges in a call graph between a caller and a called party).
All these real-world graphs are directed multigraphs, meaning that two entities may be linked by multi-edges. 
We aim to summarize and discover insightful patterns in such graphs, using a method with the desired following properties:
\begin{enumerate}
\item Robustness, to avoid detecting spurious patterns in case of noisy data.
\item Non asymptotic validity, to be able to detect reliable patterns with as few data as possible.
\item Asymptotic convergence to the true underlying distribution when it exists.
\item Parameter less
\footnote{Parameter-less means that there are no user parameter or hyper-parameter to set and that the model parame\-ters are automatically inferred during the training process.}
, with no user parameters to set.
\item Scalable, to process large graphs.
\end{enumerate}

\smallskip
In this paper, we present a novel way of analyzing and summarizing the structure of large graphs, based on piecewise constant edge density estimation.
We apply data grid models \cite{BoulleHOPR10} to graph data, where each edge is considered as a statistical unit with two variables, the source and target vertices. 
The objective is to find a correlation model between the two variables by the means of a data grid model, which in this case turns out to be a coclustering of both the source and target vertices of the graph. 
The cells resulting from the cross-product of the two clusterings summarize the edge density in the graph.
The best correlation model is selected using the MODL approach \cite{BoulleML06}, and optimized by the means of combinatorial heuristics with super-linear time complexity.
The MODL approach is an application of the Minimum Description Length (MDL) principle \cite{Rissanen78}, specialized in the following way: use of hierarchical prior uniform at each stage of the hierarchy of the model parameters, use of discrete model parameters that are data dependent and use of combinatorial optimization algorithm to find the MAP (Maximum a Posteriori) model. 

Compared with our previous work that introduced data grid models \cite{BoulleHOPR10}, we suggest a new interpretation of these models as universal density estimators and apply them to directed multigraphs, with potential loops and multi-edges, by considering these graphs as distributions of edges with unknown density. 
We also show that in our model selection approach, the finite data sample is modeled directly, with a data-dependent prior distribution of the model parameters: this provides a non-asymptotic validity of the method.
Furthermore, we demonstrate new fundamental results that prove the consistency of the approach, with an asymptotic convergence to the true underlying distribution when it exists.
Finally, we relate our method to existing approaches and present extensive comparative experiments.
Throughout the paper, all the experiments are performed on a Windows PC with Intel Xeon W3530 2.8 Ghz,
using the Khiops\
\footnote{Khiops is a general purpose data preparation and scoring tool available as a shareware at \url{http://www.khiops.com}, which implements the MODL approach described in Section~\ref{secMODLforGraphs}.}
software for the graph coclusterings based on our approach.

The rest of the paper is organized as follows.
Section~\ref{secRelatedWork} explores related work.
Section~\ref{secMODLforGraphs} reformulates the MODL method for data grid models in terms of finite data sample modeling and applies it to graphs.
Section~\ref{secConsistentEdgeDensityEstimation} introduces the problem of edge density estimation, demonstrates the asymptotically consistency of the MODL approach, and provides experimental results regarding the convergence rate.
Section~\ref{secComparativeExperiments} points out the differences between our approach and three alternative methods, using artificial datasets in controlled comparative experiments.
Section~\ref{secRealWorldExperiments} shows that our method can be used both for exploratory analysis and as a preparation step for supervised learning, using three real world datasets.
Finally, Section~\ref{secConclusion} gives a summary and suggests future work.

\section{Related Work}
\label{secRelatedWork}

Many approaches aiming at summarizing and finding structures in graphs have been proposed in the literature. 
In this section, we discuss several of these approaches and relate them to our approach.

\subsection{Graph clustering} Many approaches have been studied for the problem of graph clustering, including hierarchical clustering, divisive clustering, spectral methods, random walk (for a survey, see \cite{Schaeffer07,Fortunato10}).
To evaluate the quality of a clustering \cite{AlmeidaEtAl11} regardless of the cluster number, the modularity criterion  \cite{NewmanEtAl03} is now widely accepted in the literature and has even been treated as an objective function in clustering algorithms \cite{ClausetEtAl04,BlondelEtAl08}.
This criterion aims to obtain dense clusters where the within-cluster edge density is significantly above the expected edge density in case of random edges following the same vertex degree distribution.
Actually, not all graphs follow a \emph{cluster tendency} \cite{BezdekEtAl02}, with a structure consisting of natural clusters.
Yet, all clustering algorithms output a partition into clusters for any input graph.
While the clustering setting is relevant in many domains, our approach does not rely of such cluster tendency assumption and may have a wider range of application. 
This is illustrated experimentally in Section~\ref{secModularity}.

\subsection{Blockmodeling}
More expressive graph models aim at searching a partition of both the source and target vertices into clusters, with different types of interaction between clusters. 
The cross-product of the two partitions of vertices form a partition of the edges into \emph{blocks} or \emph{coclusters}.
This modeling approach is called blockmodeling and has been thoroughly studied for decades.
In early approaches \cite{LorrainEtAl71,ArabieEtAl78}, non-stochastic blocks are considered, with a focus on predefined types of block patterns. The blockmodel is searched either indirectly  using a (dis)similarity measure between pairs of vertices and then applying a standard clustering algorithm,
or directly by optimizing an {\it ad hoc} function measuring the fit of real blocks to the corresponding predefined types of blocks.
The limit of these approaches is that they do not cope with the stochastic nature of many real world datasets.

\subsection{Stochastic blockmodeling}
Using the framework of the exponential family, sto\-chastic blockmodels are introduced by Holland {\it et al.} \cite{HollandEtAl83}, with blocks still specified a priori.
The approach is extended by Wasserman and Anderson \cite{WassermanEtAl87} to the discovery of block structure and exploits a statistical criterion, {\it e.g.} likelihood function, optimized using the EM algorithm.
The method of Snijders and Nowicki \cite{SnijdersEtAl97} considers blockmodels where the edge probabilities depend only on the blocks to which the vertices belong. The considered models are limited to two blocks, and searched via maximum likelihood estimation using the EM algorithm for small graphs and via Bayesian Gibbs sampling for larger graphs.
The blockmodels are broadened to an arbitrary number of blocks \cite{NowickiEtAl01}, and optimized via Monte Carlo Markov Chain (MCMC) Bayesian inference.
Karrer and Newman \cite{KarrerEtAl11} propose to include the degree distribution of the vertices as a correction to the blockmodels in order to better fit real world graphs.
For a survey on recent work on stochastical blockmodeling via maximum likelihood methods, see \cite{GoldenbergEtAl10}.
In the connected approach named ``mixed membership blockmodels'' \cite{AiroldiEtAl08}, the mixture models are approximated owing to variational methods, which offer better scalability at the expense of approximating the objective function.
Non-parametric extensions of the stochastic blockmodeling approach have also been considered. In \cite{KempEtAl06}, a Dirichlet process is exploited as a prior for partitions of any size, to cluster both the source and target vertices of a directed simple graph, where the edges within each block are generated according to a Bernouilli distribution. Still, hyper-parameters are required in these approaches, such as the concentration parameter in the Dirichlet process, which influences the expected number of clusters in the non-asymptotic regime.
Our approach is closely related to non-parametric
\footnote{Here, we use the term non-parametric like in \cite{Robert97} for models where the number of parameters is not fixed and may grow with the sample size.}
stochasticl blockmodeling approaches. It differs from existing approaches on the main following points: it is not restricted to simple graphs, the model selection method exploits a MAP (maximum a posteriori) approach with an exact analytical criterion, not a Bayesian approach aiming at approximating the posterior distribution of the models, 
it is parameter-less, with no user parameter to set, and it exploits scalable optimization heuristics.
A comparative experiment with a statistical block modeling method is presented in Section~\ref{secBlockmodeling}.

\subsection{Coclustering}
A directed multigraph is fully described by its adjacency matrix, where each entry of the matrix contains the number of edges between a source vertex (in a row) and a target vertex (in a target). This is equivalent to the contingency table between two categorical variables in a dataset.
Such contingency tables can be summarized using coclustering methods.
In the applied mathematics field, the seminal work of \cite{Hartigan72} treats the problem of coclustering of a numerical matrix, by looking at a partition of the rows and columns of the matrix.
In the data mining field, in case of binary variables, this technique has been applied to the simultaneous partitioning of the instances into clusters and of the variables into groups of variables \cite{Bock79}, with methods like \cite{GovaertEtAl03} closely related to the stochastic blockmodeling approach.
Coclustering has also been applied to the domain of gene expression data \cite{ChengEtAl00,ChoEtAl04} by minimizing the sum squared residue to approximate a numerical matrix.
The method of \cite{DhillonEtAl03} optimizes a minimum loss of information to summarize a binary distance, with an application to text mining.
While most approaches require the number of row and column clusters as user parameters, the method of \cite{IencoEtAl12} is parameter-less, by directly optimizing the Goodman-Kruskal's $\tau$ measure of association between two categorical variables.
Like the method of \cite{IencoEtAl12}, our approach focuses on contingency tables containing frequency data and is parameter-less.
However, our method, being fully regularized, is both resilient to over-fitting and able to approximate the true joint distribution when it exists.

\subsection{Minimum description length based methods}
In the method of \cite{ChakrabartiEtAl04}, the MDL principle \cite{Rissanen78,Grunwald07} is employed for the inference of the whole set of blockmodel parameters, including the number of blocks, in case of directed simple graphs.
Using a two-part scheme for encoding for the model parameters and the data given the model, a parameter-less regularized criterion is obtained, inheriting from the MDL method's resistance to over-fitting.
The criterion is optimized using a greedy top-down heuristic, by adding one cluster at a time from a single-cluster initial model, and optimizing each coclustering model by moving values across clusters.
In \cite{PapadimitriouEtAl05}, the MDL method of \cite{ChakrabartiEtAl04} is extended to the case of spatial data mining.
The method of \cite{RosvallEtAl07} is dedicated to undirected simple graph, and the objective function is optimized using simulated annealing.
Following these MDL methods, several encoding schemes are explored in \cite{LangEtAl09} in the case of undirected simple graph to study their resistance to over-fitting. A fast multi-level algorithm is exploited to generate candidate partitions of the vertices of varying sizes, with a focus on the single versus multiple cluster question. The study shows that earlier approaches tend to over-fit the data with more than one cluster in case of random graphs, especially in case of skewed degree distribution of the vertices.
The early method of \cite{ChakrabartiEtAl04}, which applies to directed simple graphs, is the closest to our method. 
The main differences are that our method (1) can be applied to directed multigraph, (2) relies on an exact analytical criterion without any asymptotic approximation (such as using empirical entropies to encode the data, like in previous MDL-based methods), (3) is valid both in the non-asymptotic and asymptotic case, and (4) exploits a  bottom-up greedy optimization heuristic.
The impact of these differences both on artificial and real-world datasets is assessed in Sections~\ref{secMDLCoclustering} and \ref{secRealWorldExperiments}.

\subsection{Alternative binary matrix summarization approaches}
Other approaches have been proposed to extract patterns from binary datasets. 
For example, a tile \cite{GeertsEtAl04} is a region of a database defined by a subset of rows and columns with a high density of 1, and a collection of tiles constitutes a tiling. A tile is then closely related to one single dense cocluster, and a tiling to a coclustering, although the tiling is not a partition of the database.
In \cite{DeBieEtAl10,DeBie11}, the maximum entropy principle \cite{Jaynes03} is applied with row and columns marginals as prior information, leading to an interestingness measure of tiles and a method for extracting a tiling.
In \cite{DeBieEtAl11}, the same framework is applied for extracting multi-relational patterns: by representing multi-relational data as a K-partite graph, extracting complete connected subgraphs reduces to the problem of extracting tiles.
Coclustering has also been extended by considering a hierarchy of coclustering, where each cocluster is itself partitioned into a set of sub-coclusterings.
In \cite{PapadimitriouEtAl10}, the MDL method of \cite{ChakrabartiEtAl04} is extended to find such patterns automatically, whereas in \cite{RoyEtAl08}, the problem is treated using the Mondrian process, a multidimensional generalization of the Dirichlet process.
Tiling has also been extended to hierarchies of tiling in \cite{GionisEtAl04}.
In \cite{MiettinenEtAl08}, a binary matrix is decomposed as a product of Boolean factor matrices; extending standard matrix factorization methods, the proposed approach allows a better interpretability of the extracted patterns.
Compared with these alternative pattern extraction approaches in binary matrices, our method focuses on the problem of coclustering of a contingency matrix (or adjacency matrix of a directed multigraph), which is a matrix of counts.

\section{MODL Approach for Graphs}
\label{secMODLforGraphs}

Data grid  models \cite{BoulleHOPR10} have been introduced for the data preparation phase of the data mining process \cite{ChapmanEtAl00}, which is a key phase, both time consuming and critical for the quality of the results.
They allow one to automatically, rapidly and reliably evaluate the class conditional probability of any subset of variables in supervised learning and the joint probability in unsupervised learning. 
Data grid models are based on a partitioning of the values of each variable into intervals in the numerical case and into groups of values in the categorical case. The cross-product of the univariate partitions forms a multivariate partition of the representation space into a set of cells. This multivariate partition, called the data grid, can be interpreted as a piecewise constant non-parametric estimator of the conditional or joint probability. The best data grid is searched using a MAP approach and efficient combinatorial heuristics.
The method is non-parametric in the statistical sense, since it does not rely on the assumption that the data are drawn from a given probability distribution. 
It is also parameter-less, since all the model parameters, which number grows with the sample size, are automatically inferred without any user parameter.

\subsection{Basic Notions of Graph Theory}
\label{secGraphTheory}
A graph $G = (V, E)$ consists of a set $V$ of vertices and a set $E$ of pairs of vertices called edges.
A graph is undirected if the edges are unordered pairs of vertices, and is directed if the edges are ordered. A loop is an edge from one vertex to itself. 
A graph is simple in case of at most one edge per pair of vertices, and is a multigraph otherwise. 

Two vertices of an undirected graph are called adjacent if there is an edge connecting them. 
An edge is incident to its two vertices, called extremities. 
The degree of a vertex is the number of edges incident to it. 
In case of directed graph, the extremities of an edges are called the source and target vertices of the edge, the in-degree of a vertex $v$ is the number of edges with target $v$, and the out-degree of $v$ is the number of edges with source $v$.

Graphs can be represented by their adjacency matrix, where each cell of the matrix contains the number of edges per pair of vertices. 
The adjacency matrix of simple graphs contain only binary values, and that of undirected graphs is symmetrical. 
Figure~\ref{directed_graph} displays an example of directed simple graph.
Figure~\ref{directed_multigraph} displays a directed multigraph with self-loops, as well as it adjacency matrix and the in and out-degrees of each vertex.

\subsection{MODL Criterion for Graphs}
\label{secMODLCriterion}

\begin{figure}[!htb]
\begin{center}
\begin{minipage}[c]{.45\linewidth} \begin{center}
	\includegraphics[viewport=75 550 390 760, width=0.7\linewidth]{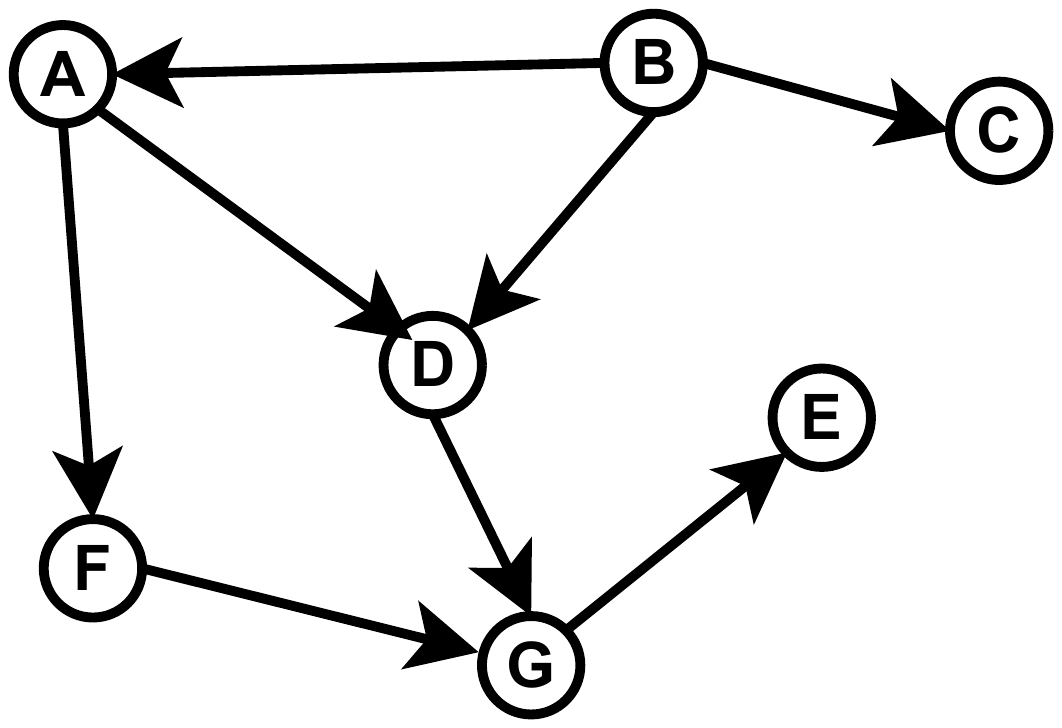}
	\quad
\end{center} \end{minipage}
\begin{minipage}[c]{.35\linewidth}
\begin{center} \begin{scriptsize} \begin{tabular}
{|c|c|} \hline
Source & Target \\ \hline
A & D \\ 
A & F \\
B & A \\ 
B & C \\ 
B & D \\ 
D & G \\ 
F & G \\ 
G & E \\ 
\hline
\end{tabular} \end{scriptsize} \end{center}
\end{minipage}
\end{center}
	\caption{Directed simple graph and its tabular representation.}
	\label{directed_graph}
\end{figure}

We reformulate the data grid approach in the context of edge density estimation in directed multigraphs.
As shown in Figure~\ref{directed_graph}, a directed graph can be represented in a tabular format with two variables, source vertex and target vertex, and one line per edge described by its two vertices. We can then apply the data grid models in the unsupervised setting to estimate the joint density between these two variables, which is the density of edges in the graph.

In this section, we formulate the approach as a modeling of a finite data sample, where the model parameters aim to summarize the edge counts in the sample.

Let $S$ and $T$ be a source and target set of vertices, with $n_S=|S|$ source vertices and $n_T=|T|$ target vertices, and let $G$ be a directed multigraph with $m$ edges from $S$ to $T$.
Given $S$, $T$ and $m$, our objective is to provide a joint description of the source and target vertices of the edges in the graph.
One simple way to describe the edges exploits the tabular format shown in Figure~\ref{directed_graph}, with the count of edges per pair (Source, Target) of vertices.
We can also summarize the location of edges at a coarser grain by introducing clusters of source vertices and clusters of target vertices, and considering the number of edges per pair of source and target cluster (cocluster).
Such a coclustering model provides a summary of the graph.
The coarsest summary is based on one single cluster of vertices with just the total number of edges, whereas the finest summary exploits one cluster per vertex, with the number of edges per pair of vertices.
Coarse grained summaries tend to be reliable, whereas fine grained summaries are more informative.
The issue is to find a trade-off between the informativeness of the summary and its reliability, on the basis of the granularity of the coclustering.

For given sets of source and target vertices $S, T$ and a given number of edges $m$ (sample size), we exploit a family of graph coclustering  models, formalized in Definition~\ref{EdgeDensityModel}.

\smallskip
\begin{definition}
\label{EdgeDensityModel}
A graph coclustering model is defined by:
\begin{itemize} 
	\item the numbers $k_S, k_T$ of source and target clusters of vertices,
	\item the partition of the $n_S$ source vertices into the $k_S$ source clusters, resulting in $n_i^S$ vertices per cluster, $1 \leq i \leq k_S$,
	\item the partition of the $n_T$ target vertices into the $k_T$ target clusters, resulting in $n_j^T$ vertices per cluster, $1 \leq j \leq k_T$,
	\item the distribution of the $m$ edges of the graph $G$ on the $k_E = k_S k_T$ coclusters with edge counts $\{m_{ij}^{ST}\}_{1 \leq i \leq k_S, 1 \leq j \leq k_T}$ per cocluster, 
	\item for each source cluster of vertices $i$, $1 \leq i \leq k_S$, the distribution of the $m_{i.}^S$ edges originating in source cluster $i$ on the $n_i^S$ vertices of the cluster, {\it i.e.} the out-degrees $\{m_{i.}\}_{1 \leq i \leq n_S}$ per source vertex,
	\item for each target cluster of vertices $j$, $1 \leq j \leq k_S$, the distribution of the $m_{.j}^T$ edges terminating in target cluster $j$ on the $n_j^T$ vertices of the cluster, {\it i.e.} the in-degrees $\{m_{.j}\}_{1 \leq j \leq n_T}$ per target vertex.
\end{itemize}
\end{definition}

\begin{table}[!htb]
\caption{Notation}
\begin{center}
\begin{tabular}{|l|l|}
\hline
	$S, T$ & source and target vertex sets \\
	$n_S=|S|$ & number of source vertices \\
	$n_T=|T|$ & number of target vertices \\
	$G$ & directed multigraph with edges from $S$ to $T$ \\
	$m$ & number of edges in $G$ \\
\hline
  $M$ & graph coclustering model for given $S, T, m$ \\
	$k_S, k_T$ & number of clusters of source and target vertices \\
	$k_E = k_S k_T$ & number of coclusters of edges \\
	$m_{ij}^{ST}$ & number of edges for cocluster $(i,j)$ \\
	$m_{i.}$ & number of edges for source vertex $i$ (out-degrees) \\
	$m_{.j}$ & number of edges for target vertex $j$ (in-degrees) \\
  & \\
	$n_i^S$ & number of vertices in source cluster $i$ \\
	$n_j^T$ & number of vertices in target cluster $j$ \\
	$m_{i.}^S$ & number of edges originating in source cluster $i$ \\
	$m_{.j}^T$ & number of edges terminating in target cluster $j$ \\
\hline
	$m_{ij}$ & number of edges for pair $(i,j)$ of vertices \\ 
\hline
\end{tabular}
\end{center}
\label{tab:notation}
\end{table}

\begin{figure}[!htb]
\begin{center}
\begin{minipage}[c]{.34\linewidth} \begin{center}
	\includegraphics[viewport=75 550 390 760, width=0.9\linewidth]{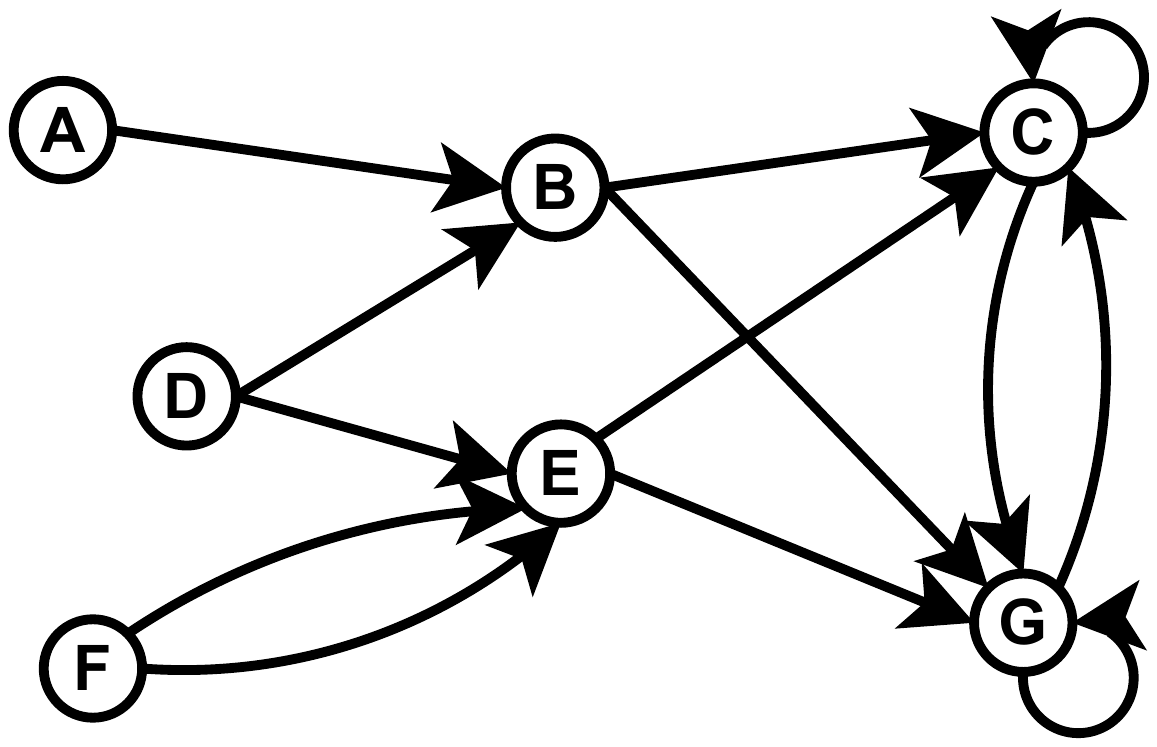}
\end{center} \end{minipage}
\begin{minipage}[c]{.64\linewidth} \begin{center} \begin{scriptsize} \begin{tabular}
{r |c|c|c|c|c|c|c|l} 
\multicolumn{1}{c}{} & \multicolumn{1}{c}{A} & \multicolumn{1}{c}{B} & \multicolumn{1}{c}{C} &
\multicolumn{1}{c}{D} & \multicolumn{1}{c}{E} & \multicolumn{1}{c}{F} & \multicolumn{1}{c}{G} & $\sum{}$ \\ \cline{2-8}
A & 0 & 1 & 0 & 0 & 0 & 0 & 0 & 1 \\ \cline{2-8}
B & 0 & 0 & 1 & 0 & 0 & 0 & 1 & 2 \\ \cline{2-8}
C & 0 & 0 & 1 & 0 & 0 & 0 & 1 & 2 \\ \cline{2-8}
D & 0 & 1 & 0 & 0 & 1 & 0 & 0 & 2 \\ \cline{2-8}
E & 0 & 0 & 1 & 0 & 0 & 0 & 1 & 2 \\ \cline{2-8}
F & 0 & 0 & 0 & 0 & 2 & 0 & 0 & 2 \\ \cline{2-8}
G & 0 & 0 & 1 & 0 & 0 & 0 & 1 & 2 \\ \cline{2-8}
\multicolumn{1}{r}{$\sum{}$} & \multicolumn{1}{c}{0} & \multicolumn{1}{c}{2} & \multicolumn{1}{c}{4} &
\multicolumn{1}{c}{0} & \multicolumn{1}{c}{3} & \multicolumn{1}{c}{0} & \multicolumn{1}{c}{4} & 13 \\ 
\end{tabular} \end{scriptsize} \end{center} \end{minipage}
\end{center}
	\caption{Directed multigraph and its adjacency matrix. The numbers $m_{ij}$ in the adjacency matrix are the numbers of edges for each pair of vertices (for example, two edges from F to E). The sums $m_{i.}$ on the right column are the out-degrees of the vertices, and the sums $m_{.j}$ on the bottom line are the in-degrees of the vertices. The total number of edges is on the bottom right corner of the adjacency matrix.}
	\label{directed_multigraph}
\end{figure}

\begin{figure}[!htb]
\begin{center}
\begin{minipage}[c]{.34\linewidth} 
\begin{flushleft}
	\includegraphics[viewport=75 550 390 760, width=0.85\linewidth]{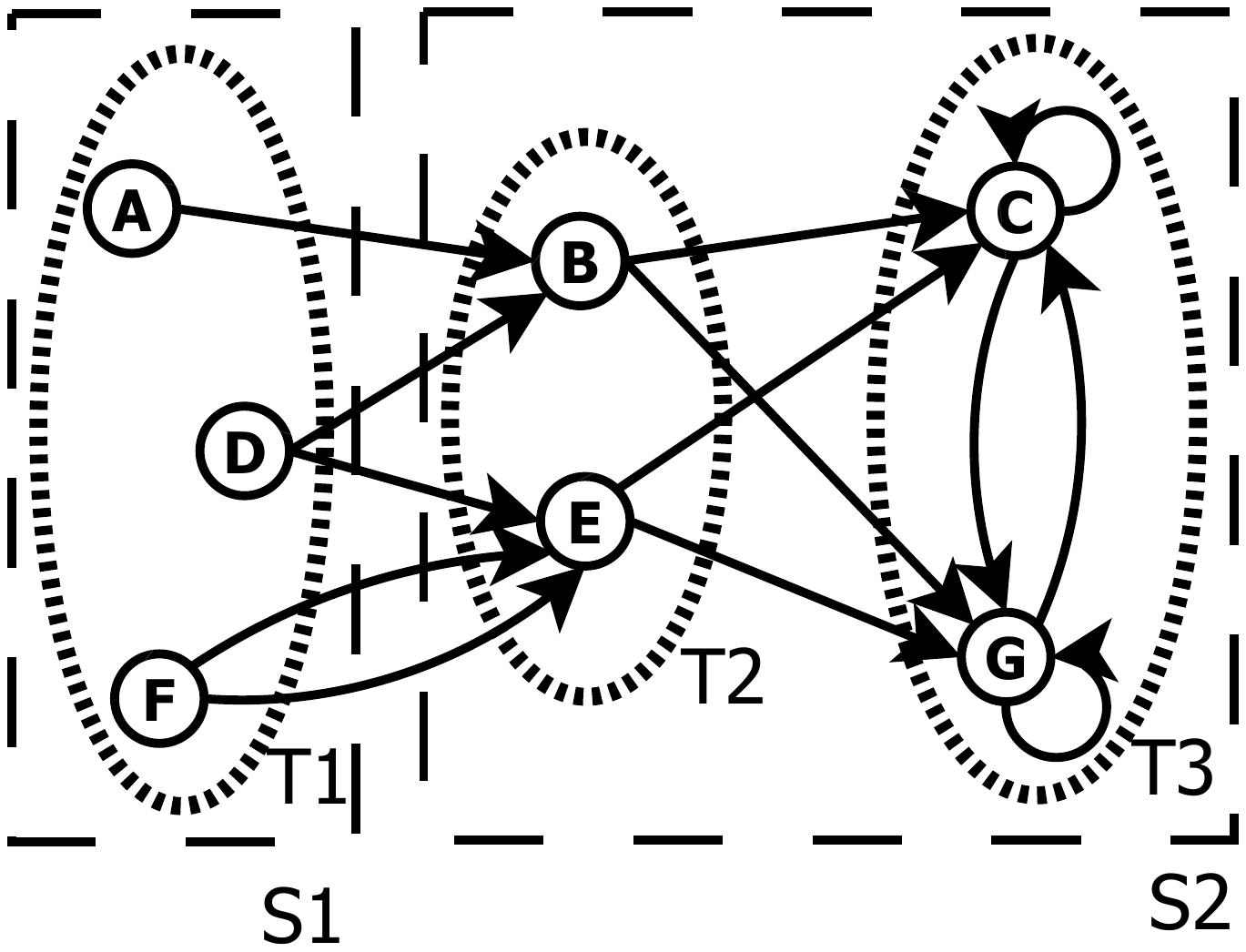}
\end{flushleft} 
\end{minipage}
\begin{minipage}[c]{.64\linewidth} \begin{center} \begin{scriptsize} \begin{tabular}
{r|c|c|c||c|c||c|c|l} 
\multicolumn{1}{c}{} & \multicolumn{1}{c}{A} & \multicolumn{1}{c}{D} & \multicolumn{1}{c}{F} &
\multicolumn{1}{c}{B} & \multicolumn{1}{c}{E} & \multicolumn{1}{c}{C} & \multicolumn{1}{c}{G} & $\sum{}$ \\ \hhline{~|---||--||--|~}
A & 0 & 0 & 0 & 1 & 0 & 0 & 0 & 1 \\ \hhline{~|---||--||--|~}
D & 0 & 0 & 0 & 1 & 1 & 0 & 0 & 2 \\ \hhline{~|---||--||--|~}
F & 0 & 0 & 0 & 0 & 2 & 0 & 0 & 2 \\ \hhline{~:===::==::==:~}
B & 0 & 0 & 0 & 0 & 0 & 1 & 1 & 2 \\ \hhline{~|---||--||--|~}
E & 0 & 0 & 0 & 0 & 0 & 1 & 1 & 2 \\ \hhline{~|---||--||--|~}
C & 0 & 0 & 0 & 0 & 0 & 1 & 1 & 2 \\ \hhline{~|---||--||--|~}
G & 0 & 0 & 0 & 0 & 0 & 1 & 1 & 2 \\ \hhline{~|---||--||--|~}
\multicolumn{1}{r}{$\sum{}$} & \multicolumn{1}{c}{0} & \multicolumn{1}{c}{0} & \multicolumn{1}{c}{0} &
\multicolumn{1}{c}{2} & \multicolumn{1}{c}{3} & \multicolumn{1}{c}{4} & \multicolumn{1}{c}{4} & 13 \\ 
\end{tabular} \end{scriptsize} \end{center} \end{minipage} 
\begin{minipage}[c]{.25\linewidth} \begin{center} \begin{scriptsize} \begin{tabular}
{r|c||c||c|l} 
\multicolumn{1}{c}{} & \multicolumn{1}{c}{} & \multicolumn{1}{c}{} & \multicolumn{1}{c}{} &  \\
\multicolumn{1}{c}{} & \multicolumn{1}{c}{T1} & \multicolumn{1}{c}{T2} & \multicolumn{1}{c}{T3} & $\sum{}$ \\ \hhline{~|-||-||-|~} 
S1 & 0 & 5 & 0 & 5 \\ \hhline{~:=::=::=:~} 
S2 & 0 & 0 & 8 & 8 \\ \hhline{~|-||-||-|~} 
\multicolumn{1}{r}{$\sum{}$} & \multicolumn{1}{c}{0} & \multicolumn{1}{c}{5} & \multicolumn{1}{c}{8} & 13 \\ 
\end{tabular} \end{scriptsize} \end{center} \end{minipage}
\end{center}
	\caption{Directed multigraph with two source and three target clusters. The adjacency matrix of the graph (reorganized by clusters) is presented on the top-left, and that of the clustered graph at the bottom. The numbers $m_{ij}^{ST}$ in the clustered adjacency matrix are the numbers of edges for each cocluster (for example, 5 edges from S1 to T2).}
	\label{directed_clustered_multigraph}
\end{figure}

This notation, summarized in Table~\ref{tab:notation}, is illustrated in Figure~\ref{directed_multigraph}, where a directed multigraph is displayed with its adjacency matrix.
A clustered version of this graph is presented in Figure~\ref{directed_clustered_multigraph}, which results in a coclustering of its adjacency matrix.

We assume that the numbers of edges $m$ and of source and target vertices $n_S$ and $n_T$ are known in advance and we aim to model the $m$ edges of $G$ between these two sets of vertices.
This setting is general enough to account for directed graphs, bipartite graphs and undirected graph, where each edge comes twice with the two directions.

The family of models introduced in Definition~\ref{EdgeDensityModel} is completely defined by the numbers $k_S$ and $k_T$ of clusters,  the partitions of vertices into clusters, the edge counts in each coclusters 
$ \{m_{ij}^{ST}\}_{1 \leq i \leq k_S, 1 \leq j \leq k_T},$	 
and the out- and in-degree of each vertex
$\{m_{i.}\}_{1 \leq i \leq n_S}, \{m_{.j}\}_{1 \leq j \leq n_T}.$

The numbers of vertices per cluster $n_i^S$ and $n_j^T$ are derived from the specification of the partitions of vertices into clusters: they do not belong to the model parameters. Similarly, the numbers of edges originating or terminating in each cluster can be deduced by adding the frequencies of coclusters, according to $m_{i.}^S = \sum_{j=1}^{k_T} {m_{ij}^{ST}}$ and $m_{.j}^T = \sum_{i=1}^{k_S} {m_{ij}^{ST}}$.

\smallskip 
In order to select the best model, we apply a MAP approach.
We suggest the prior distribution of the model parameters described in Definition~\ref{EdgeDensityPrior}, by applying the following modeling choices: use of discrete rather than real-valued distributions, of the ``natural'' hierarchy of the model parameters and choice of uniform distributions at each level of the hierarchy, to be as uninformative as possible.

\smallskip 
\begin{definition} 
\label{EdgeDensityPrior}
The prior for the parameters of a graph coclustering model are chosen hierarchically and uniformly at each level: 
\begin{itemize} 
\item	the numbers of clusters $k_S$ and $k_T$ are independent from each other, and uniformly distributed between $1$ and $n_S$ for the source vertices, between $1$ and $n_T$ for the target vertices,
\item for a given number $k_S$ of source clusters, every partition of the $n_S$ vertices into $k_S$ clusters is equiprobable,
\item for a given number $k_T$ of target clusters, every partition of the $n_T$ vertices into $k_T$ clusters is equiprobable,
\item for a model of size $(k_S, k_T)$, every distribution of the $m$ edges on the $k_E=k_S k_T$ coclusters is equiprobable,
\item	for a given cluster of source (resp. target) vertices, every distribution of the edges originating (resp. terminating) in the cluster on the vertices of the cluster is equiprobable.
\end{itemize}
\end{definition}
\smallskip 

\smallskip
Finally, we now introduce the notion of \emph{consistency of a graph coclustering model with a graph sample} in Definition~\ref{GraphConsistency}.

\begin{definition} 
\label{GraphConsistency}
For given sets of vertices $S, T$ and number of edge $m$, a graph coclustering model $M$ is \emph{consistent} with a graph sample $G$ if and only if the edge counts $\{m_{ij}^{ST}\}$, $\{m_{i.}\}$, $\{m_{.j}\}$ in the model are exactly the same as the related empirical edges counts $\{{M}_{ij}^{ST}\}$, $\{{M}_{i.}\}$, $\{{M}_{.j}\}$ in the graph sample.
\end{definition}

A model which is not consistent with the data cannot generate the data and obtains a null posterior probability.
We now focus on the posterior probability of consistent models to obtain the evaluation criterion given in Theorem~\ref{EdgeDensityTheorem}~\cite{BoulleHOPR10}.

\smallskip 
\begin{theorem}
\label{EdgeDensityTheorem}
The negative log of the posterior probability of a graph coclustering model $M$ consistent with a graph sample $G$, distributed according to a uniform hierarchical prior, is given by
{\allowdisplaybreaks
\begin{subequations} \label{E:EDC}
\begin{align}
c(M) =& \log n_S + \log n_T  \label{E:EDC1} \\ 
&+ \log B(n_S, k_S) + \log B(n_T, k_T)  \label{E:EDC2} \\
& + \log \binom {m + k_E - 1} {k_E - 1}  \label{E:EDC3} \\
&  + \sum_{i=1}^{k_S} {\log \binom {m_{i.}^S + n_i^S - 1} {n_i^S - 1}}  \label{E:EDC4} \\
&  + \sum_{j=1}^{k_T} {\log \binom {m_{.j}^T + n_j^T - 1} {n_j^T - 1}}  \label{E:EDC5} \\
& + \log m! - \sum_{i = 1}^{k_S} {\sum_{j = 1}^{k_T} {\log m_{ij}^{ST}!}}  \label{E:EDC6} \\
& + \sum_{i = 1}^{k_S} {\log m_{i.}^S!} 
  - \sum_{i = 1}^{n_S} {\log m_{i.}!}  \label{E:EDC7} \\
&  + \sum_{j = 1}^{k_T} {\log m_{.j}^T!}
  - \sum_{j = 1}^{n_T} {\log m_{.j}!}  \label{E:EDC8} 
\end{align}
\end{subequations}}
\end{theorem}
\smallskip 

$B(n,k)$ is the number of divisions of $n$ elements into $k$ subsets (with potentially empty subsets). When $n=k$, $B(n,k)$ is the Bell number. In the general case, $B(n,k)$ can be written as $B(n,k) = \sum_{i=1}^k {S(n,i)}$, where $S(n,i)$ is the Stirling number of the second kind \cite{AbramowitzEtAl70}, which stands for the number of ways of partitioning a set of $n$ elements into $i$ nonempty subsets.

Mainly, the evaluation criterion of Theorem~\ref{EdgeDensityTheorem} relies on counting the number of possibilities for the model parameters and for the data given the model.
In Equation~\ref{E:EDC}, line (\ref{E:EDC1})
\footnote{
For the choice of an integer parameter $k$ uniformly distributed between 1 and $n$, we get $p(k) = \frac {1} {n}$, leading to $-\log k = \log n$.
}
relates to the prior distribution of the cluster numbers $k_S$ and $k_T$ 
and line (\ref{E:EDC2})
\footnote{
For a partition of $n$ elements into $k$ subsets chosen unifomly among $B(n,k)$ possibilities, we get $\log 
B(n,k)$ terms in the criterion.
}
to the specification of the partition of the source (resp. target) vertices into  clusters. 
These terms are the same as in the case of the MODL supervised univariate value grouping method \cite{BoulleJMLR05}.
Line (\ref{E:EDC3})
\footnote{
The number of positive integer parameters $(n_1, \ldots, n_k)$ with $\sum_{i=1}^k {n_i}=n$ is $\binom {n + k - 1} {k - 1}$.
}
represents the specification of the parameters of the distribution of the $m$ edges on the $k_E$ coclusters, followed in line (\ref{E:EDC4}) (resp. line (\ref{E:EDC5})) by the specification of the distribution of the edges originating (resp. terminating) in each cluster on the vertices of the cluster.
Line (\ref{E:EDC6})
\footnote{
The number of ordered partitions of $n$ elements into $k$ subsets of sizes $(n_1, \ldots, n_k)$ is given by the multionomial coefficient.
}
stands for the likelihood of the distribution of the edges on the coclusters, by the means of a multinomial coefficient.
Finally, line (\ref{E:EDC7}) (resp. line (\ref{E:EDC8})) corresponds to the likelihood of the distribution of the edges originating (resp. terminating) in each cluster on the vertices of the cluster.

As negative log of probabilities are code lengths, our model selection technique is similar to a practical minimum description length principle \cite{Rissanen78,Grunwald07} with two-part MDL code.
Our method is valid non-asymptotically, since it directly encodes the edge counts of the data sample. The inferred MAP model is necessarily consistent with the data sample: it provides a valid summary of the edge counts in the graph sample, but without any asymptotic guarantee w.r.t. the underlying edge probability distribution.
In Section~\ref{secConsistentEdgeDensityEstimation}, we study the asymptotic behavior of the approach as the number of edges in the data sample goes to infinity, and demonstrate that it can be interpreted as an edge density estimator with asymptotic convergence to the true edge distribution when it exists.

\subsection{Optimization Algorithm}
\label{secOptimizationAlgorithm}

Graph coclustering models are no other than data grid models \cite{BoulleHOPR10} applied to the case of joint density estimation of the source and target vertices of the edges.
The space of data grid models is so large that straightforward algorithms almost surely fail to obtain good solutions within a practicable computational time. Sophisticated heuristics are described in \cite{BoulleHOPR10} to optimize the criterion $c(M)$. They finely exploit the sparseness of the adjacency matrix of the graph and the additivity of the  criterion, and allow a deep search in the model space with $\mathcal{O}(m)$ memory complexity and $\mathcal{O}(m \sqrt m \log m)$ time complexity.

In this section, we give an overview of the optimization algorithms which are fully detailed in \cite{BoulleHOPR10}, and rephrase them using the graph terminology.
The optimization of a data grid is a combinatorial problem.
The number of possible partitions of $n$ vertices is equal to the Bell number $B(n)=\frac{1}{e} \sum_{k=1}^{\infty} {\frac {k^n} {k!}}$. 
Even with very simple models having only two clusters of source and target vertices, the number of models involves $2^{n_S + n_T}$ coclusterings of the vertices.
An exhaustive search through the whole space of models is unrealistic. We describe in Algorithm~\ref{GBUM} a greedy bottom up merge heuristic (GBUM) which optimizes the model criterion $c(M)$.
The method starts with a fine grained model, with few vertices per source or target cluster, up to the maximum model $M_{\mbox{\footnotesize Max}}$ with one vertex per source or target cluster.
It considers all the merges between clusters (independently for the source and target sets of vertices), and performs the best merge if the criterion decreases after the merge. 
The process is reiterated until no further merge decreases the criterion.

\begin{algorithm} [!htbp]
\caption{Greedy Bottom Up Merge heuristic (GBUM)}
\label{GBUM}
\begin{algorithmic} [1]
  \REQUIRE $M$ \COMMENT {Initial solution}
  \ENSURE $M^*, c(M^*) \leq c(M)$ \COMMENT {Final solution with improved cost}
  \STATE $M^* \leftarrow M$
  \WHILE {improved solution}
    \STATE $M^{\prime} \leftarrow M^*$
    \FORALL {$merge$ between two source or target clusters}
    \STATE \COMMENT {Consider $merge$ for model $M^*$}
    \STATE $M^+ \leftarrow M^* + merge$  
    \IF {$c(M^+) < c(M^{\prime})$}
    	\STATE $M^{\prime} \leftarrow M^+$
    \ENDIF
    \ENDFOR
    \IF {$c(M^{\prime}) < c(M^*)$}
    	\STATE $M^* \leftarrow M^{\prime}$ \COMMENT {Improved solution}
    \ENDIF
  \ENDWHILE
\end{algorithmic}
\end{algorithm}

Each evaluation of the criterion for a model requires $\mathcal{O}(n^2)$ time, since the initial model contains up to $n_S n_T$ coclusters (see Equation~(\ref{E:EDC})) in the case of the maximal model $M_{\mbox{\footnotesize Max}}$. 
Each step of the algorithm relies on $\mathcal{O}(n^2)$ evaluations of merges of clusters of vertices, and there are at most $\mathcal{O}(n)$ steps, since the model becomes equal to the null model $M_\emptyset$ (one single cluster) once all the possible merges have been performed. 
Overall, the time complexity of the algorithm is $\mathcal{O}(n^5)$ using a straightforward implementation of the algorithm.
However, the method can be optimized in $\mathcal{O}(m \sqrt m \log m)$ time. The optimized algorithm mainly exploits the sparseness of the data, the additivity of the criterion and starts from non-maximal models with pre and post-optimization heuristics.
\begin{itemize}
	\item 
Large graph are often sparse, with far less edges than in complete graphs.
Although a model may contain $\mathcal{O}(n^2)$ coclusters, at most $m$ coclusters are non empty.
Since the contribution of empty coclusters is null in the criterion \ref{E:EDC}, each evaluation of a data grid can be performed in $\mathcal{O}(m)$ time owing to specific algorithmic data structures (mainly, sparse representation with fast access to edges via hash-indexes).
	\item 
The additivity of the criterion means that it can be decomposed on the hierarchy of the components of the  models: extremity (sources vs target variable), cluster of vertices, cocluster. Using this additivity property, all the merges between adjacent clusters can be evaluated in $\mathcal{O}(m)$ time. Furthermore, when the best merge is performed, the only impacted merges that need to be reevaluated  for the next optimization step are the merges that share edges with the best merge. Since the graph is potentially sparse, the number of reevaluations of models is small on average.
	\item 
Finally, the algorithm starts from initial fine grained solutions containing at most $\mathcal{O}(\sqrt m)$ clusters.
Specific pre-processing and post-processing heuristics are exploited to locally improve the initial and final solutions of Algorithm~\ref{GBUM} by moving vertices across clusters. 
The post-optimization algorithms are applied alternatively to the source and target vertex variables, for a frozen partition of the other variable.
This allows one to keep a $\mathcal{O}(m)$ memory complexity and to bound the time complexity by $\mathcal{O}(m \sqrt m \log m)$. 
\end{itemize}
Sophisticated algorithmic data structures are necessary to exploit these optimization principles and guarantee a time complexity of $\mathcal{O}(m \sqrt m \log m)$ for initial solutions exploiting at most $\mathcal{O}(\sqrt m)$ clusters of vertices.

The optimized version of the greedy heuristic is time efficient, but it may fall into a local optimum. 
This problem is tackled using the variable neighborhood search (VNS) meta-heuristic \cite{HansenEtAl01}, which mainly benefits from multiple runs of the algorithms with different random initial solutions.
The main heuristic described in Algorithm~\ref{GBUM}, with its guaranteed time complexity, is used to find a good solution as quickly as possible. The VNS meta-heuristic is exploited to perform anytime optimization: the more you optimize, the better the solution. To favor quality over speed, the meta-heuristic default setting is to perform at least 10 rounds of the main optimization heuristic before stopping.

The optimization algorithms summarized above have been extensively evaluated \cite{BoulleHOPR10}, using a large variety of artificial datasets, where the true data distribution is known. 
Overall, the method is both resilient to noise and able to detect complex fine grained patterns. It is able to approximate any data distribution, provided that there are enough instances in the train data sample.

\section{Consistency of the Approach for Edge Density Estimation}
\label{secConsistentEdgeDensityEstimation}

In this section, we interpret the models presented in Section~\ref{secMODLforGraphs} as edge density estimators, demonstrate that the MODL approach converges asymptotically to the true edge density distribution when it exists. We also provide experimental results regarding the convergence rate of the approach.

\subsection{Edge Density Estimation}
\label{secEdgeDensityEstimation}

In our approach, we consider the graphs as generative models, where the statistical units are the edges with two variables per edge, the source and target vertices of the edge.
Whereas most blockmodeling approaches deal with simple graphs, focusing on their topology with at most one edge per pair of vertices, we regard graphs as statistical distributions of directed edges, with potential loops and multi-edges.
A graph edge density model for a set of $n$ vertices is entirely defined by a set of probability parameters $\{p_{ij}\}_{1 \leq i \leq n, 1 \leq j \leq n}$, where $p_{ij}$ stands for the probability of each independent and identically distributed (i.i.d) edge having source vertex $i$ and target vertex $j$.
Given these settings, a graph $G$ containing $m$ edges is treated as a sample of size $m$ drawn from the edge distribution.
Therefore, large samples tend be produce complete graphs from a pure topological point of view, but with varying edge densities taking into account the generative model.

This edge density model applies to much real world graph data. In web log analysis, it seems natural to consider a bipartite graph, with users as source vertices, web pages as target vertices and edges representing web navigation. A sample graph corresponds to an extract of web log data, with the popular pages much more seen than the others.
In a phone call network, each edge represents one phone call from a caller vertex to a called vertex, so that two vertices can be connected by multi-edges. Collecting the phone calls during a given time period corresponds to a sample of a directed multigraph, where the potential communities correspond to subgraphs with high multi-edge density.
The case of undirected graphs can be treated with symmetrical edge probabilities and a pair of directed edges per undirected edge
\footnote{
Although our approach can deal with undirected graphs using both directions, such graphs would benefit from a specialized prior (potentially very different) and from simpler optimization algorithms (no need for alternate optimization of the partitions, for example).}
.

Given this random graph generative model, the problem is to estimate the edge densities in the graph from a finite data sample.
Estimating the $n^2$ edge probability parameters $p_{ij}$ from a sample of size $m$ is not an easy task, especially in the case of sparse graphs.

\subsection{MODL Approach for Edge Density Estimation}
\label{secEdgeDensityApproach}

In the following, we propose a new interpretation of our approach described in Section~\ref{secMODLforGraphs} and show how it reduces to a finite sample modeling, which asymptotically converges to an estimation of the edge density parameters.

Given a graph coclustering model $M$ as defined in Section~\ref{secMODLforGraphs}, let us introduce the following notation for the probability of the edges in a coclustered random graph:
\begin{itemize} 
	\item $\{p_{\kappa \lambda}^{ST}\}_{1 \leq \kappa \leq k_S, 1 \leq \lambda \leq k_T}$: probability distribution of the edges falling in each cocluster $(\kappa, \lambda)$
	\item $\{p_{\kappa, i.}\}_{k_S(i)=\kappa}$: probability distribution of the out-degrees of the vertices $i$ of the source cluster $\kappa$
	\item $\{p_{\lambda,.j}\}_{k_T(j)=\lambda}$: probability distribution of the in-degrees of the vertices $j$ of the target cluster $\lambda$
\end{itemize}

Using this notation, the probability parameters of a coclustered random graph can be empirically estimated from the edge count parameters in a model $M$ according to:
\begin{equation}
\label{parameterEstimation}
p_{\kappa \lambda}^{ST} = \frac {m_{\kappa \lambda}^{ST}} {m}, \qquad
p_{\kappa, i.} =  \frac {m_{i.}} {m_{\kappa.}^S}, \qquad
p_{\lambda,.j} =  \frac {m_{.j}} {m_{.\lambda}^T}.
\end{equation}

This is a piece-wise constant modeling of the edge density with respect to the coclusters, constrained by the distributions of the in and out-degrees of the vertices in each cluster.
Assuming the independence between the source and target vertices of the edges inside each cocluster, we get the following estimation of the edge densities:
\begin{equation}
\label{edgeDensityestimation}
p_{i j} = p_{\kappa \lambda}^{ST} p_{\kappa, i.} p_{\lambda,.j}
                 = \frac {m_{\kappa j}^{ST}} {m} \frac {m_{i.}} {m_{\kappa.}^S} \frac {m_{.j}} {m_{.\lambda}^T},
\end{equation}
where $(\kappa ,\lambda)$ is the cocluster containing the edge $(i, j)$.

For the null model $M_{\emptyset}$ with one single cluster ($\kappa =\lambda =1$), we have
\begin{equation}
\label{parameterEstimationNull}
p_{11}^{ST} = 1, \;
p_{1, i.} =  \frac {m_{i.}} {m}, \;
p_{1,.j} =  \frac {m_{.j}} {m}, \;
p_{i j} = \frac {m_{i.}} {m} \frac {m_{.j}} {m},
\end{equation}
which means that the joint probability distribution $p_{ij}$ is the product of the two independent marginal distributions of the in and out-degrees of the vertices.

For the maximal model $M_{\mbox{\footnotesize Max}}$ with one cluster per vertex, we have
\begin{equation}
\label{parameterEstimationMaximal}
p_{ij}^{ST} = \frac {m_{ij}} {m}, \;
p_{i, i.} = 1, \;
p_{j,.j} = 1, \;
p_{i j} = \frac {m_{i j}} {m},
\end{equation}
which means that the joint probability distribution $p_{ij}$ of the edges is directly estimated by the model parameters.

\subsection{Asymptotic Convergence of the MODL Approach}
\label{secMODLAsymtoticConsistency}

The family of coclustered random graphs is very expressive and can theoretically approximate any edge distribution provided that there is sufficient data.
The problem is to select the best model given the data.

The MODL approach aims to model directly the finite data sample, and exploits a discrete model space of the edge counts in the sample graph.
Working on a set of parameters of finite size allows one to define a ``natural'' hierarchical prior with uniform distribution at each level of the hierarchy, as in Definition~\ref{EdgeDensityPrior}, and reduces to counting in this discrete model space.
This data-dependent modeling technique leads to the criterion of Equation~\ref{E:EDC}, which can be interpreted as the exact posterior probability of the sample graph given the model (Bayesian interpretation), or the exact code length of the model parameters and edges given the model (two-part MDL interpretation \cite{Rissanen78,Grunwald07}).
Therefore, the criterion does not rely on empirical estimation of continuous-valued parameters (such as probabilities or entropies), which are valid only asymptotically.
We now study whether for given sets of source and target vertices, this exact finite data sample modeling asymptotically converges towards the true edge density when it exists, as the edge number goes to infinity.

Let us first recall some concepts from information theory.
The Shannon entropy $H(X)$ \cite{CoverEtAl91} of a discrete random variable $X$ with probability distribution function $p$ is defined as:
\begin{equation}
\label{entropy}
H(X) = -\sum_{x \in X} {p(x) \log p(x)}.
\end{equation}
The mutual information of two random variables is a quantity which measures the mutual dependence of the two variables \cite{CoverEtAl91}; it vanishes if and only if they are independent.
For two discrete variables  $X$ and $Y$, the mutual information is defined as:
\begin{equation}
\label{mutualInformation}
I(X;Y) = \sum_{x \in X} {\sum_{y \in Y} {p(x,y) \log \frac {p(x,y)} {p(x) p(y)}}},
\end{equation}
where $p(x,y)$ is the joint probability distribution function of $X$ and $Y$, and $p(x)$ and $p(y)$ are the marginal probability distribution functions of $X$ and $Y$ respectively.

Let us consider edges as statistical instances, with two vertex variables $V_S$ and $V_T$ having $n_S$ and $n_T$ values, and two vertex cluster variables $V_S^M$ and $V_T^M$ having $k_S$ and $k_T$ values for a given coclustering model $M$.

We present in Theorem~\ref{EdgeDensityCriterionAsymptotics} an asymptotic approximation of the evaluation criterion $c(M)$ introduced in Equation~\ref{E:EDC}.

\smallskip
\begin{theorem}
\label{EdgeDensityCriterionAsymptotics}
The MODL evaluation criterion (Equation~\ref{E:EDC}) for a  graph coclustring model $M$ is asymptotically equal to $m$ times the entropy of the source and target vertex variables minus the mutual entropy of the variables grouped.
\begin{equation}
\label{EdgeDensityCriterionAsymptoticsFormula}
c(M) = m \left( H(V_S) + H(V_T) - I(V_S^M; V_T^M) \right) + \mathcal{O}(\log m).
\end{equation}
\end{theorem}

\begin{proof}
See Appendix.
\end{proof}
\smallskip

As the criterion has to be minimized, this means that the method aims to select a coclustering model which maximizes the mutual information between the two vertex cluster variables. 
Since the mutual information of two variables is not other than the Kullback-Leibler divergence \cite{CoverEtAl91} between the joint probability distribution of two variables and their independent joint distribution, this means that the best selected coclustering tends to highlight contrasts between the two variables, being as far as possible from their independent joint distribution.

We now present an important result in Theorem~\ref{EdgeDensityCriterionLimit}, which shows that the MODL approach asymptotically converges towards the estimation of the true edge distribution, which is the joint distribution of the source and target vertex variables.
Although the modeling technique is data-dependent (regarding the model space and the prior on the model parameters) and aims to model exactly the data sample with a discrete distribution of the sample edges on the vertices, not the true edge continuous-valued probability distribution, this theorem demonstrates the consistency of the approach.

\smallskip
\begin{theorem}
\label{EdgeDensityCriterionLimit}
The MODL approach for selecting a graph coclustering model $M$ asymptotically converges towards the true edge distribution, and the criterion for the best model $M_{\mbox{\footnotesize Best}}$ converges to $m$ times the entropy of the edge variable, which is the joint entropy of the source and target vertices variables.
\begin{equation}
\label{EdgeDensityCriterionLimitFormula}
\lim_{m \rightarrow \infty} \frac {c(M_{\mbox{\footnotesize Best}})} {m} = H(V_S, V_T).
\end{equation}
\end{theorem}

\begin{proof}
See Appendix.
\end{proof}
\smallskip

As a corollary of Theorem~\ref{EdgeDensityCriterionLimit}, Theorem~\ref{EdgeDensityCriterionLimitMI} states that the MODL approach allows one to estimate the mutual information between the source and target vertices variables.

\smallskip
\begin{theorem}
\label{EdgeDensityCriterionLimitMI}
The MODL approach for selecting a graph coclustering model $M$ asymptotically converges towards the true edge distribution, and the criterion for the null model minus the criterion for the best model $M_{\mbox{\footnotesize Best}}$ converges to $m$ times the mutual entropy of the source and target vertices variables.
\begin{equation}
\label{EdgeDensityCriterionLimitFormulaMI}
\lim_{m \rightarrow \infty} \frac {c(M_{\emptyset})-c(M_{\mbox{\footnotesize Best}})} {m} = I(V_S; V_T).
\end{equation}
\end{theorem}
\smallskip

Let us recall that the mutual information $I(V_S; V_T)$ is null in case of independent source and target vertices, such as for Erd\H{o}s-R{\'e}nyi random graphs \cite{ErdosEtAl76}.
Theorem~\ref{EdgeDensityCriterionLimitMI} shows that for such graphs, the best selected model will be asymptotically the same as the null model (which actually represents the case of independence). 
Since the MODL approach is regularized, with prior terms in criterion $c(M)$ that grows with the granularity of the clusters, we expect the approach to select the null model in case of independence, even in the non-asymptotic case.
This expected behavior is confirmed experimentally in Section~\ref{secComparativeExperiments}.

\subsection{Experimental Convergence Rate of the MODL Approach}
\label{secConvergenceRate}

We have shown that although the MODL approach aims to model the data sample directly, it asymptotically converges towards the true edge density.
The assumption behind the MODL approach is that the non-parametric edge density estimation will benefit from fine tuned finite data-dependent model space and prior, so as to converge as fast and reliably as possible.

This convergence rate is hard to analyze theoretically in the non-parametric setting, without any assumption regarding the true edge density.
For example, in the simple case of a cluster-based graph, the adjacency matrix is block-diagonal and most of the edge probabilities are null. In this case, few parameters need to be estimated and the convergence is fast.
In this section, we chose a more difficult sample graph where the distribution of the edge probabilities is rather smooth with no cluster-based structure, unbalanced and never null, and present an experimental study of the convergence rate of the approach.

Let us introduce \emph{circular random graphs} as directed multigraphs, where the $n$ vertices lie equidistant on the unit circle at positions $(x_i = \cos \frac {2 \pi i} {n}, y_i = \sin \frac {2 \pi i} {n})$. The Euclidian distance between two vertices $i$ and $j$ being $d_{ij} = \sqrt {(x_i-x_j)^2 + (y_i-y_j)^2}$, which we extend by continuity to $d(i, i) = \frac {2} {n}$ for self-loops, we define the probability of having an edge between two vertices in inverse proportion of their distance, according to:
\begin{equation}
\label{CircularRandomGraphEdgeDensity}
\forall i, j, \; \ p_{ij} = \frac { 1/d_{ij}}  {\sum_{\mu, \gamma} {1/d_{\mu \gamma}}}.
\end{equation}

In such a circular random graph, the largest edge probabilities (related to self-loops with $d = \frac {2} {n}$) are $n$ times larger than the smallest ones (related to pairs of vertices on a diameter of the circle, with $d=2$), with a continuous decrease of edge probabilities in inverse proportion to the distance of their extremities.

In our experiment, we chose a circular random graph with $n=100$ vertices and randomly generate edges from sample size varying from 100 to $10^6$.
For each sample size, we run the MODL algorithm and collect both the number of clusters and the mutual information estimated according to the data grid edge probability estimator (see Equation~\ref{edgeDensityestimation}) and to the MODL criterion (see Equation~\ref{EdgeDensityCriterionLimitFormulaMI} from Theorem~\ref{EdgeDensityCriterionLimitMI}).
For comparison purpose, we also report the true mutual information (known exactly for this artificial dataset), as well as its empirical estimation (using $p_{ij} = \frac {m_{ij}} {m}$) and according to the Laplace estimator ($p_{ij} = \frac {m_{ij}+1} {m+n^2}$).

\begin{figure}[!tbp]
\begin{center} \begin{tabular}{c}
	\includegraphics[viewport=20 65 750 480, width=0.48\linewidth]{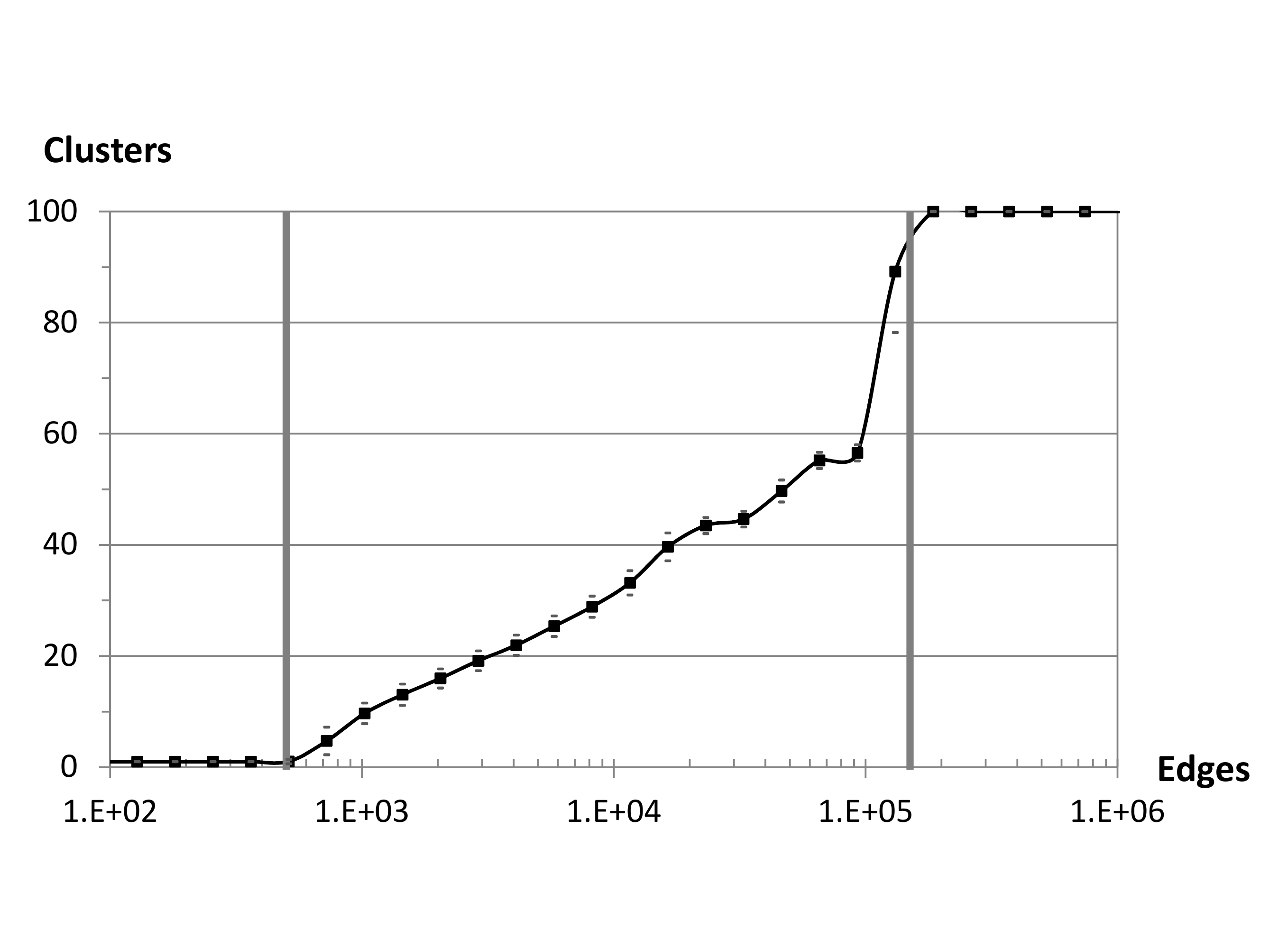}
	\,
	\includegraphics[viewport=20 65 750 480, width=0.48\linewidth]{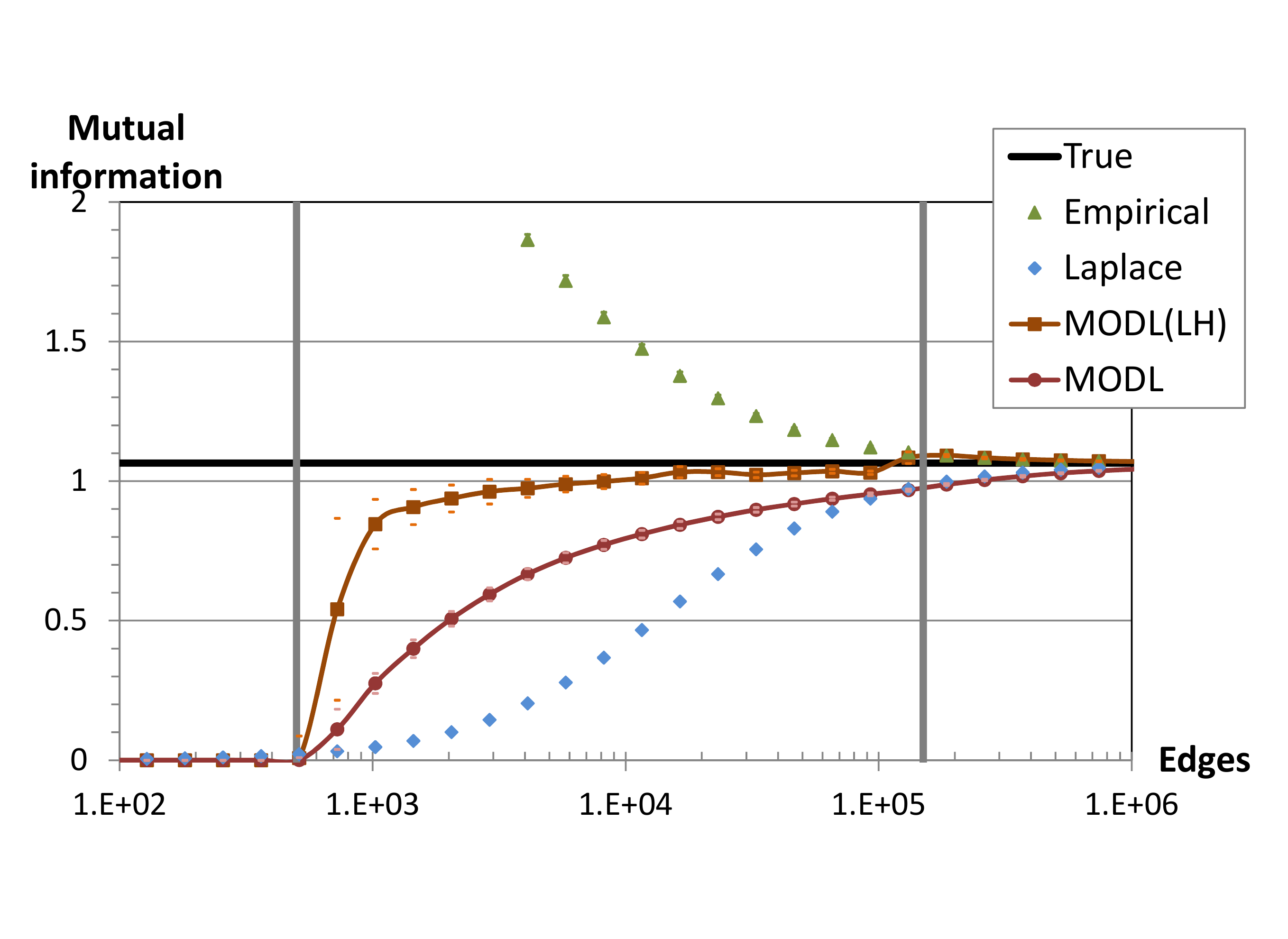} 
	\end{tabular} \end{center}
	\caption{Convergence of the MODL approach on the random circular graph with 100 vertices: number of clusters  and difference between the estimated and true mutual information $I(V_S; V_T)$ per edge number in the graph sample.}
	\label{randomCircularConvergence}
\end{figure} 

The experiment is repeated 100 times, so as to estimate both the mean and the standard deviation of the collected measures, presented in Figure~\ref{randomCircularConvergence}.
The empirical estimator tends to over-fit the data (the mutual information is largely overestimated, especially for small sample sizes), whereas the Laplace estimator tends to under-fit the data.
In accordance with Theorem~\ref{EdgeDensityCriterionLimitMI}, the MODL criterion of Equation~\ref{EdgeDensityCriterionLimitFormulaMI} converges towards the true mutual information.
The MODL(LH) criterion (likelihood terms only, without prior terms as in Equation~\ref{EdgeDensityCriterionLimitFormulaMI}) related to the best selected model exhibits a much faster convergence rate, while not over-fitting the data.

Figure~\ref{randomCircularConvergence} shows three phases in the convergence of the MODL approach.
In the first phase (\emph{stability phase}), the number of edges is not sufficient to reliably estimate the edge probabilities, and the approach evaluates the random graph with one single cluster as being the most probable.
In the second phase (\emph{non-parametric estimation phase}), the method identifies regularities in the graph by building clusters and approximating the true mutual information, with an increased accuracy as the sample size grows.
In the third phase (\emph{classical estimation phase}), the method has built one cluster per vertex and estimates all the $n^2$ edges probabilities simultaneously according to a classical empirical estimation of a set of multinomial parameters: the accuracy of the estimation ``classically'' increases with the sample size.
In this example, the non-parametric estimation phase starts when the number of available edges is about 500, about $\frac {1} {20}$ of the number of edge probability parameters, and has converged at about 150,000 edges, fifteen times the number of edge probability parameters.
It is noteworthy that in most large real world graphs, the method will run in the non-parametric estimation phase.

Overall, this experiment shows that the method is reliable, quickly discovers true regularities in the graph as soon as there is sufficient data, and is able to approximate complex edge density distributions with a fast convergence rate.

\section{Comparative Experiment on Artificial Datasets}
\label{secComparativeExperiments}

In this section, with exploit artificial datasets, where the true edge distribution is known, to compare our approach to alternative methods.
We first relate our method to a graph clustering method, then to a statistical blockmodeling method, and finally to a parameter-less coclustering method based on MDL. 

\subsection{Comparison with Graph Clustering}
\label{secModularity}

This section, intended to be of a tutorial nature, points out the difference between graph clustering and our method, which exploits a coclustering approach.
We first recall the modularity criterion which is widely used for graph clustering methods, then illustrate the difference between the approaches using artificial datasets.

\subsubsection{Modularity Criterion for Graph Clustering Methods}
\label{Modularity}

The goal of community detection is to partition a network into clusters of vertices with high edge density, with the vertices  belonging to different clusters being sparsely connected.
To evaluate the quality of a partition, the modularity $Q$ \cite{NewmanEtAl03} is a widely used criterion in recent community detection methods.
The modularity measures the density of edges inside clusters as compared to the one expected in case of independence of the vertices.

Given a graph $G$ with $n$ vertices and $m$ edges, let $m_{ij}$ be an element of the adjacency matrix of the graph. 
$m_{ij} = 1$ if vertices $i$ and $j$ are connected by an edge, $m_{ij} = 0$ otherwise.
The degree of a vertex $i$ is defined by the number of edges incident upon it. 
In the case of undirected graphs, the ouput and input degree of a vertice are equal.
Using the notation of Section~\ref{secMODLCriterion}, we have

\begin{equation}
\label{vertexDegree}
m_{i.} = m_{.i} = \sum_j {m_{ij}} = \sum_j {m_{ji}}\;.
\end{equation}

An undirected graph with $m_U$ edges corresponds to a symmetrical directed graph with $m = 2 m_U$ edges.
Assuming that the vertex degrees are respected, the probability of a random edge between vertices $i$ and $j$ is $m_{i.} m_{.j} / m^2$.
The modularity $Q$ is defined as
\begin{equation}
\label{modularityCriterion}
Q = \frac {1} {m} \sum_{ij} {\left( m_{ij} - \frac {m_{i.} m_{.j}} {m} \right)} \delta(k_S(i), k_T(j)),
\end{equation}
where $k_S(i)=k_T(i)$ is the index of the cluster to which vertex $i$ is assigned, the $\delta$-function $\delta(x,y)$ is 1 if $x=y$ and 0 otherwise and $m= \sum_{ij}{m_{ij}}$ is twice the number of (undirected) edges.
The modularity takes its values between -1 and 1 and has positive values when the clusters have more internal edges that the expected edge number if connections where made at random, with the same vertex degrees.
The value of this criterion is 0 in the two extreme cases of one single cluster and of as many clusters as vertices.
The modularity criterion has two appealing properties: it is well founded for the discovery of clusters with a density higher than the expected density when the extremities of the edges are independent, and it does not require any parameter, such as the number of clusters.

\medskip
Modularity has been used to evaluate the quality of partitions for a large variety  of methods such as hierarchical clustering, spectral clustering, random walks (see Section~\ref{secIntroduction}), but also as an objective function to optimize.
In this section, we compare our approach with the state of the art Louvain method \cite{BlondelEtAl08}, which is very fast and builds high quality partitions (measured by the modularity criterion).

\subsubsection{Artificial Graph Family}

We introduce a family of artificial graphs consisting in four clusters of ten vertices, named $A, B, C, D$. 
For two-dimensional depiction purpose, we consider the case of undirected simple graphs, with at most one edge per pair of vertices and no loops, and control the proportion of potential edges per cocluster, that is per pair of clusters of vertices.
This is illustrated in Figure~\ref{four_tiny_cluster}, where the four clusters of vertices are drawn on circles for better readability. For example, choosing a proportion $p=20\%$ for the edges of $(A,B)$ means that 20\% of the potential edges with one extremity in $A$ and the other one in $B$ (among $100=10*10$ edges) are in the graph.
In the rest of the section, we illustrate the difference between graph clustering and coclustering using two graph patterns: quasi-cliques and cocliques.

\subsubsection{Quasi-cliques}

\begin{figure}[!htbp]
\begin{center} \begin{small} \begin{tabular}{cccc}
\includegraphics[viewport=1 1 490 490, width=0.18\linewidth]{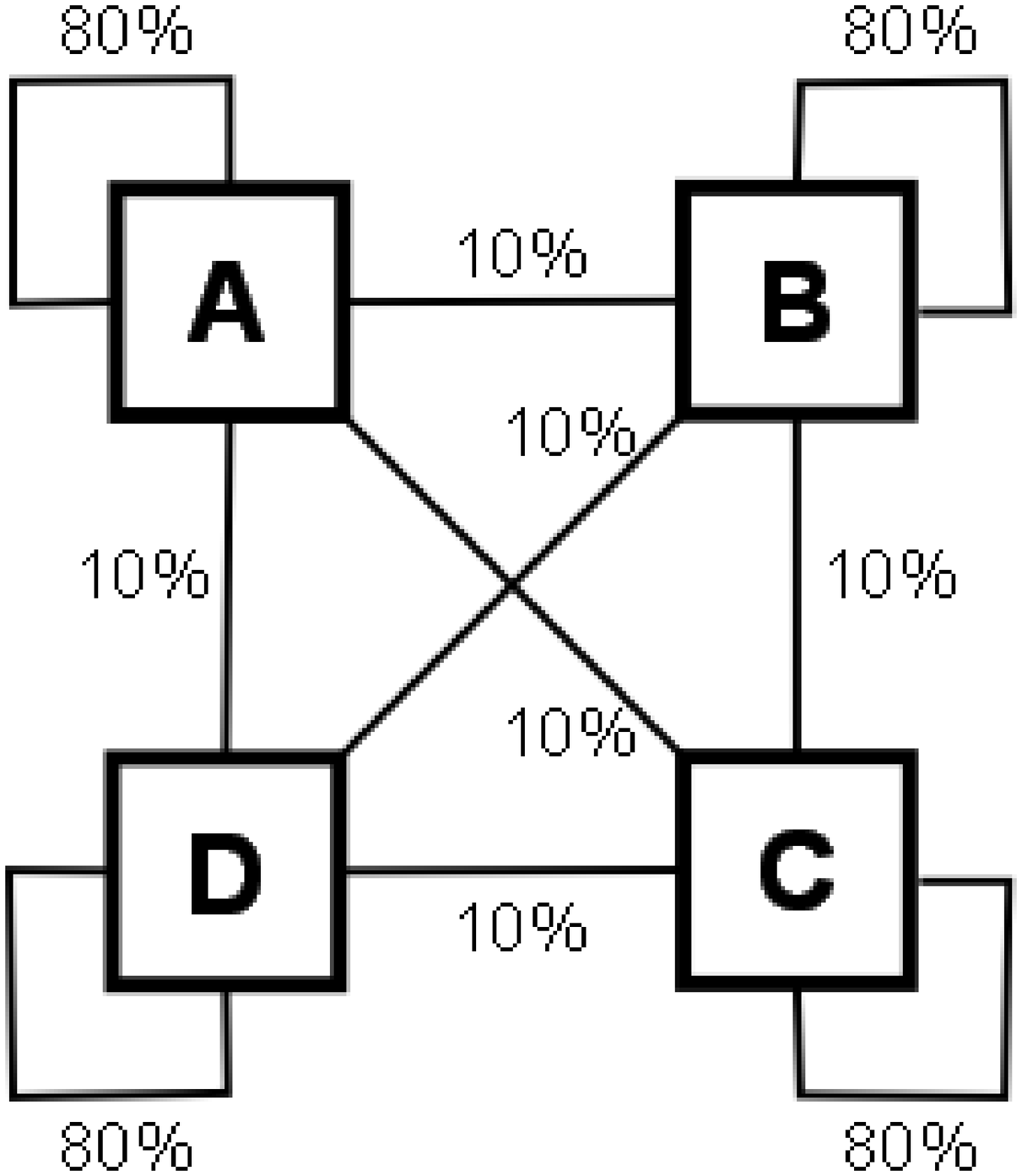} &
\includegraphics[viewport=115 230 490 610, width=0.20\linewidth]{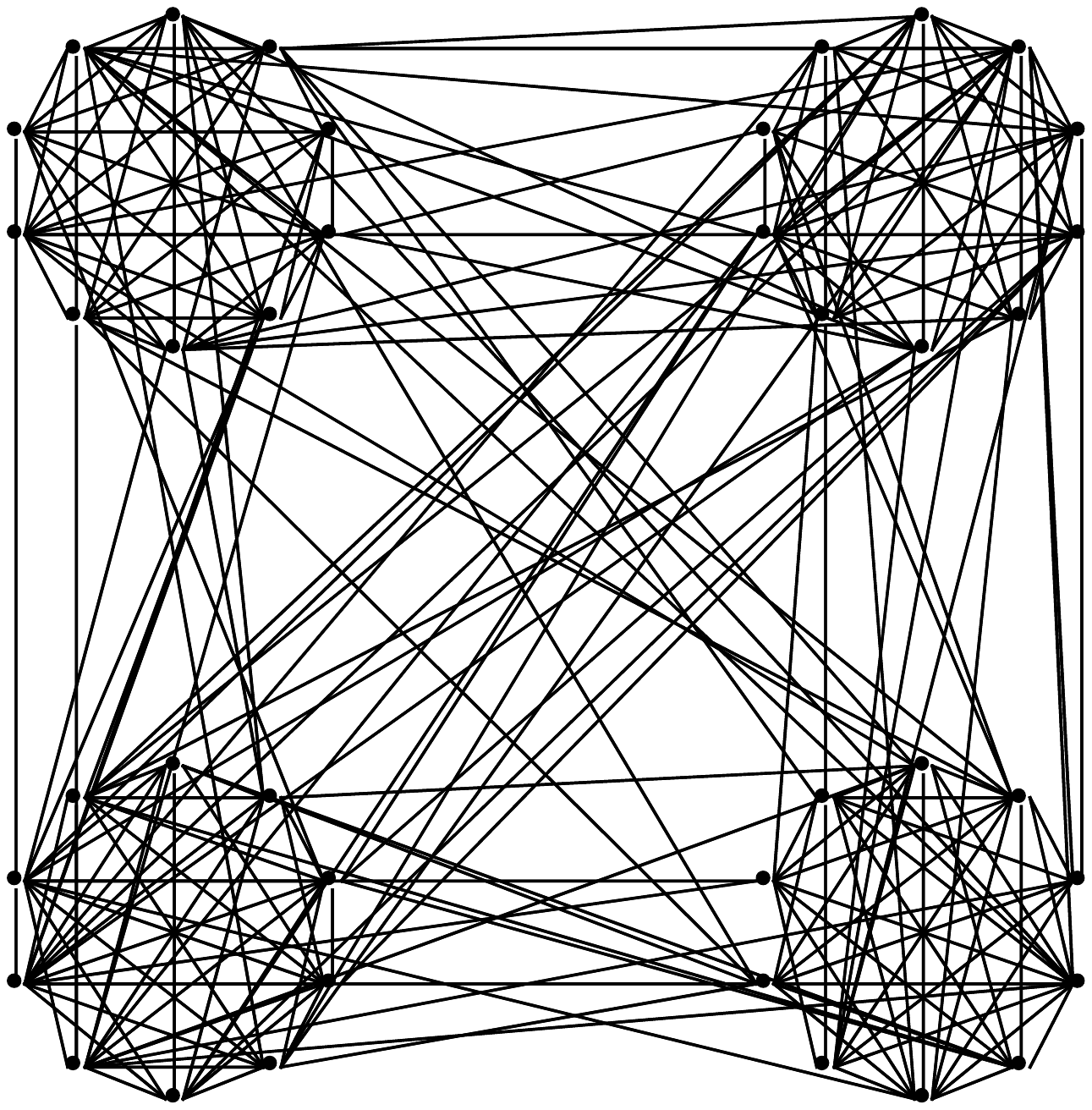} &
\includegraphics[viewport=115 230 490 610, width=0.20\linewidth]{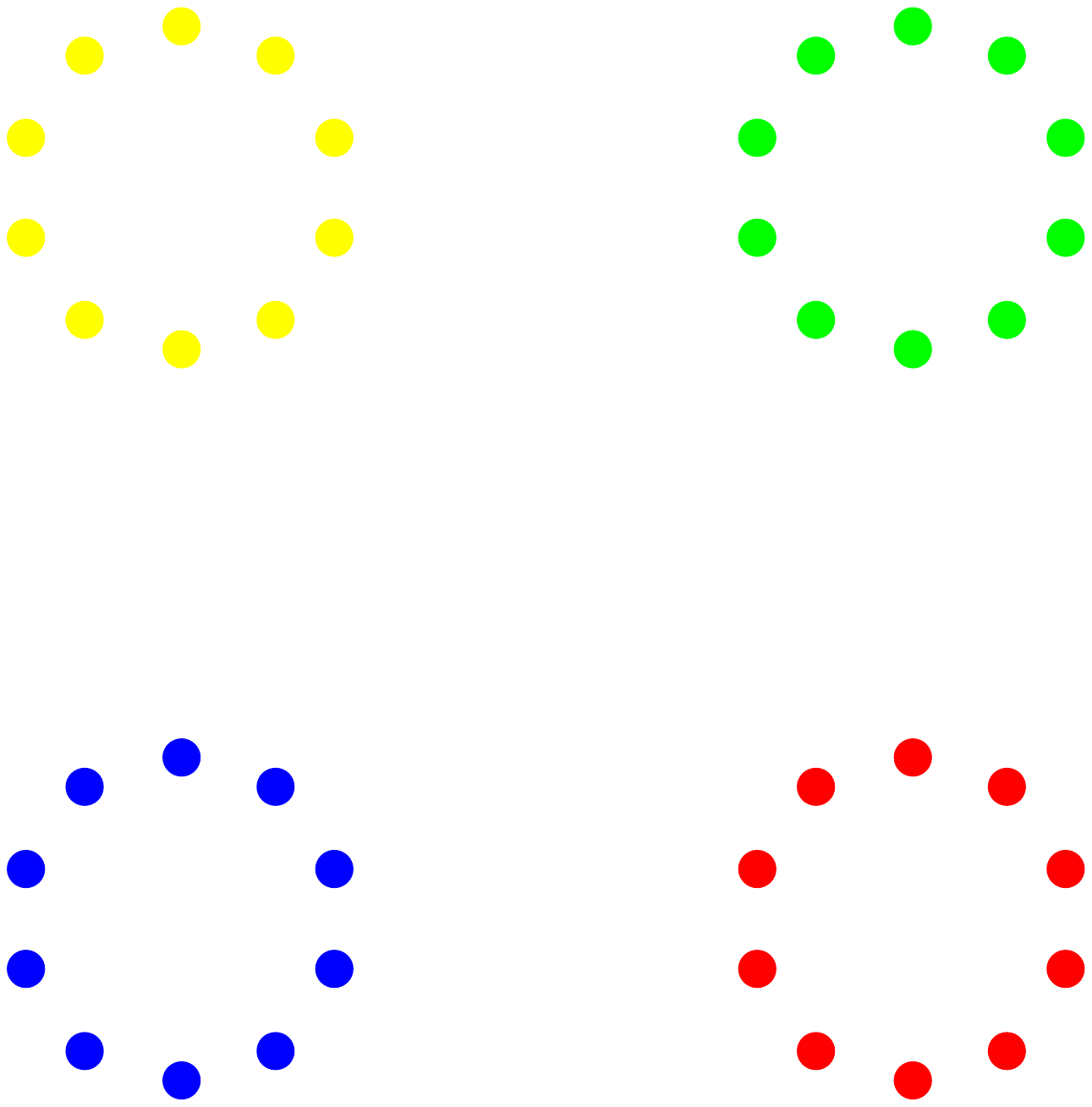} &
\includegraphics[viewport=115 230 490 610, width=0.20\linewidth]{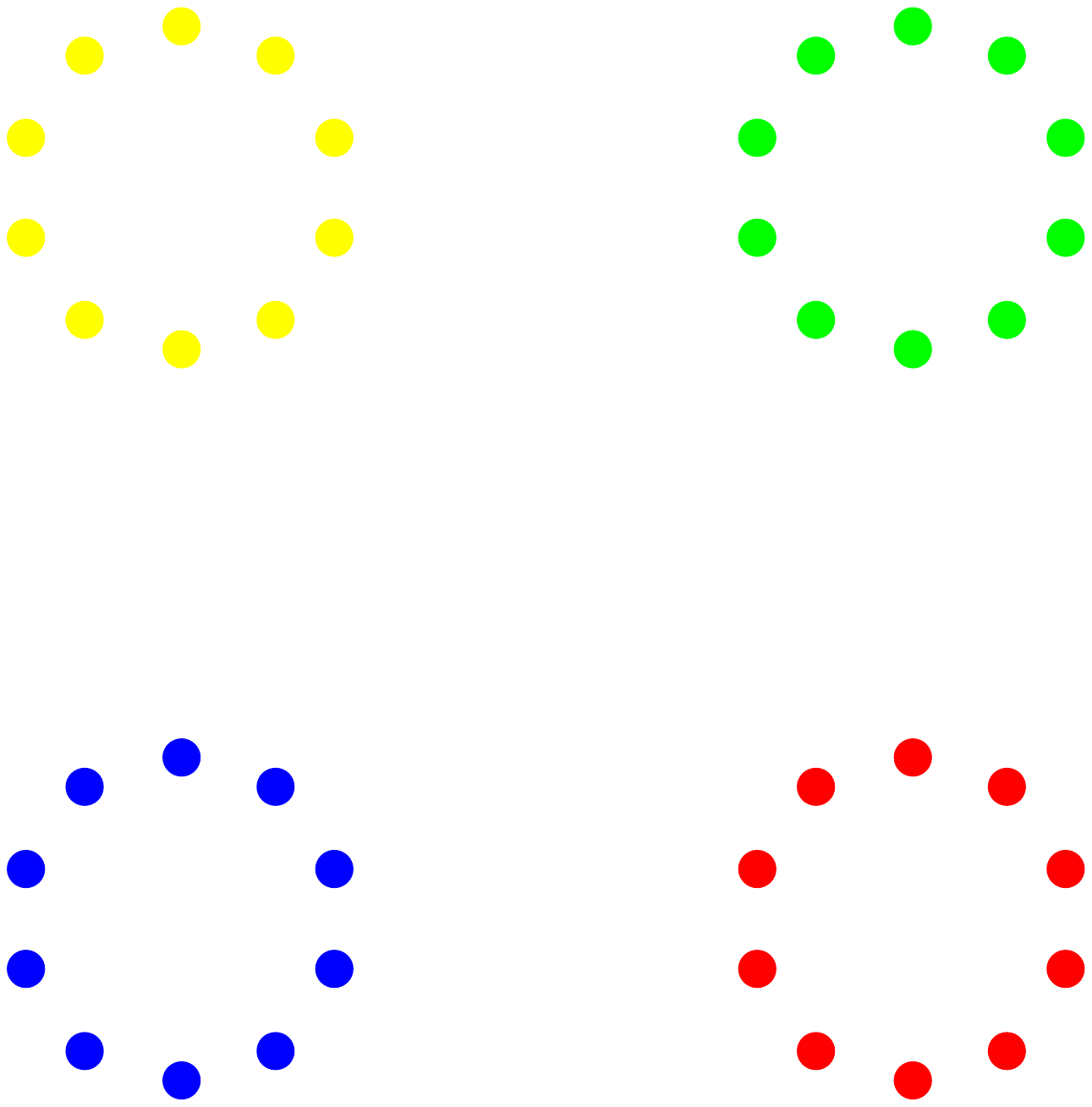} \\
Distribution & Example & Coclustering & Graph clustering
\end{tabular} \end{small} \end{center}
	\caption{Artificial graph: quasi-cliques.}
	\label{four_tiny_cluster}
\end{figure}

Figure~\ref{four_tiny_cluster} presents a classical pattern consisting of four dense clusters, with an intra-cluster density of $80\%$ and an inter-cluster density of $10\%$.
The parameters of the edges distribution are shown on the left, then an example of a graph generated according to this distribution, and on the right the clusters retrieved using our approach and the modularity-based approach, with a different color per cluster.
The cluster based and coclustering methods (modularity method of \cite{BlondelEtAl08} and our approach) obtain the same result, with a correct identification of the four clusters related to $A, B, C, D$ (with $Q=0.409$).
In the case of a graph that can be decomposed into dense clusters, the two approaches exhibit the same behavior.

\subsubsection{Cocliques}

\begin{figure}[!htbp]
\begin{center} \begin{small} \begin{tabular}{cccc}
\includegraphics[viewport=1 1 490 490, width=0.18\linewidth]{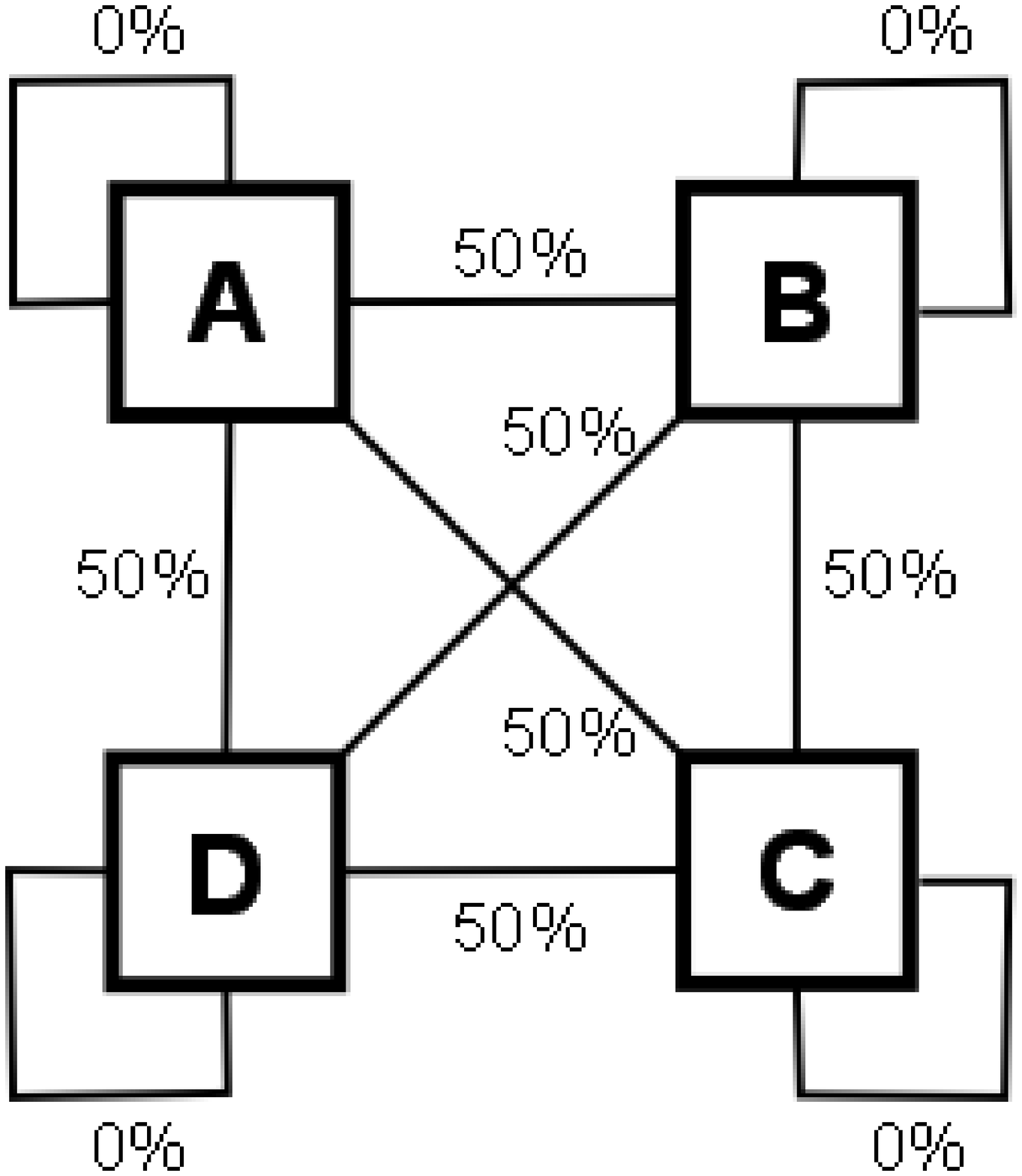} &
\includegraphics[viewport=115 230 490 610, width=0.20\linewidth]{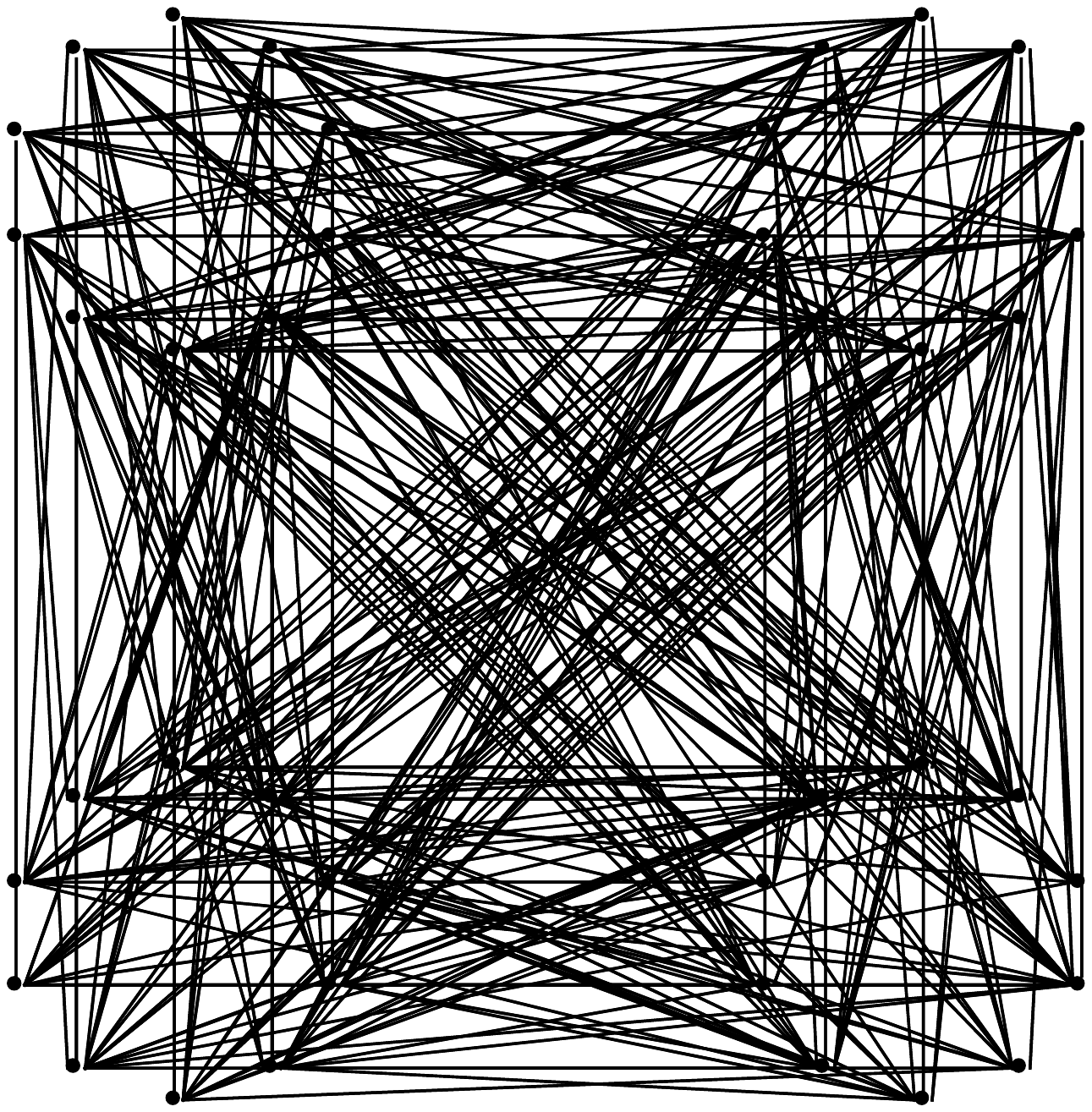} &
\includegraphics[viewport=115 230 490 610, width=0.20\linewidth]{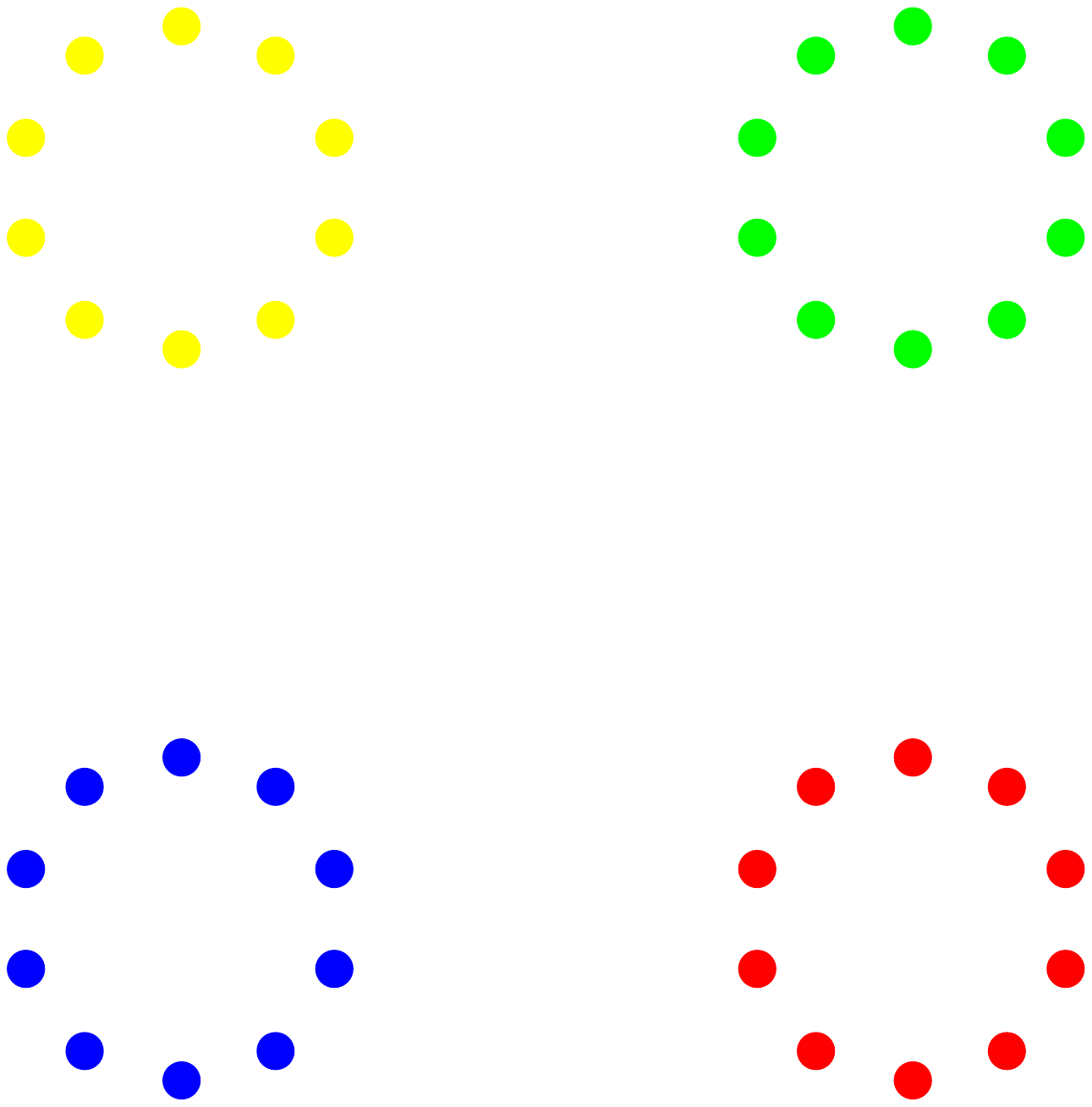} &
\includegraphics[viewport=115 230 490 610, width=0.20\linewidth]{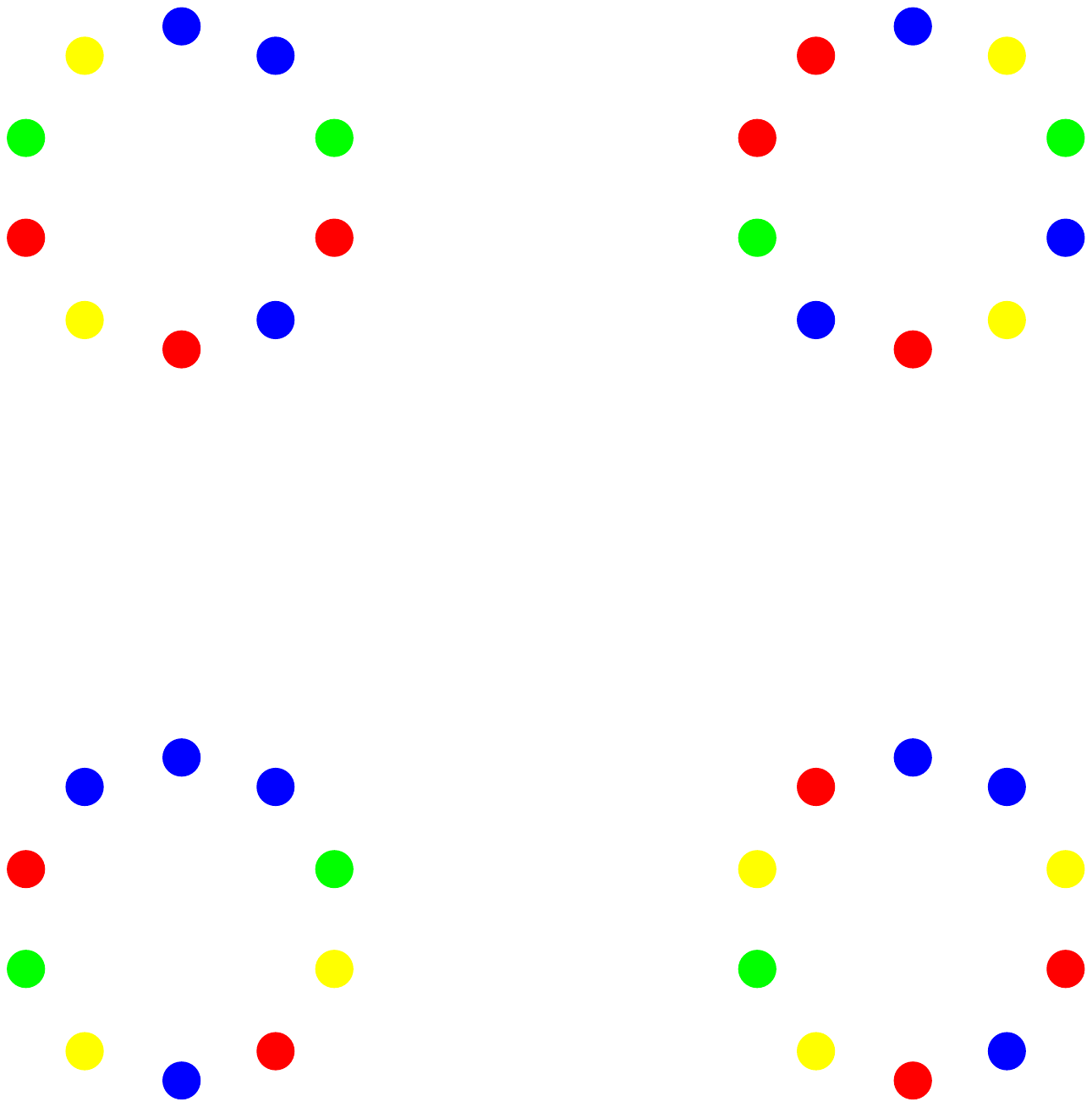} \\
Distribution & Example & Coclustering & Graph clustering
\end{tabular} \end{small} \end{center}
	\caption{Artificial graph: cocliques.}
	\label{four_tiny_empty}
\end{figure}

Figure~\ref{four_tiny_empty} presents a pattern consisting of four cocliques, which are subgraphs with no inner edge, and an edge density across cocliques of $50\%$.
In this example, the intra-cluster density is far below the average density.
Actually, not all graphs have a structure consisting of natural clusters. 
Yet, all clustering algorithms output a partition into clusters for any input graph, and the cluster based algorithm builds dubious clusters in this case. Our approach correctly retrieves the four empty cocliques. In this case, the modularity is negative (-0.251), which reflects the fact that the ratio between observed and expected edge density is far below 1.

\subsection{Comparison with Stochastic Blockmodeling}
\label{secBlockmodeling}

In this section, we focus on the blockmodeling approach and report comparative experiments on two kinds of artificial graphs: blockmodel graphs and random graphs.

\subsubsection{Blockmodel Graphs}

Our method is a piecewise constant non-parametric edge density estimator in directed multigraphs.
One of the closest existing approach is the statistical blockmodeling one, and we used the StOCNET software for comparison purpose.
StOCNET \cite{Stocnet06} is a software system for the advanced statistical analysis of social networks, which implements the stochastic blockmodeling method named BLOCKS of Nowicki and Snijders \cite{NowickiEtAl01}, which applies to simple graphs.
Although StOCNET requires the number of blocks as an input parameter, the tool can compute the blockmodeling structure for numbers of blocks between 2 and 8. 
The fit of the block structure is evaluated using the \emph{clarity} of the block structure \cite{NowickiEtAl01}.
The authors suggest to keep the model with the best fit (smallest clarity) to select the best number of blocks.

We evaluate the ability of the approach to retrieve the correct block structure using a graph consisting of 100 vertices with three clusters: 30 vertices in cluster A, 40 in cluster B and 30 in cluster C. The distribution of the edges across the clusters is summarized in Table~\ref{tableArtificialBM3}, where the source and target vertices of each edge are uniformly distributed within each cluster.
For each sample size 100, 200, ..., 1,000 edges, we generate one hundred datasets according to this edge distribution and run both the StOCNET software and our method with their default settings.
The computation time of StOCNET is on average 210 seconds, while our methods takes on average 0.2 seconds.

\begin{table}[!htb]
\label{tableArtificialBM3}
\begin{center} \begin{footnotesize} 
\caption{Artificial blockmodel distribution with three clusters.}
\begin{tabular}{l|c|c|c|c} 
\multicolumn{1}{l}{} & 
\multicolumn{1}{c}{A} &
\multicolumn{1}{c}{B} &
\multicolumn{1}{c}{C} &
 $\sum{}$ \\ \hhline{~|-|-|-|~}  
A	& 30\%	&  0\%	&  0\%	& 30\% \\ \hhline{~|-|-|-|~} 
B	&  0\%	& 10\%	& 30\%	& 40\% \\ \hhline{~|-|-|-|~} 
C	&  0\%	& 30\%	&  0\%	& 30\% \\ \hhline{~|-|-|-|~} 
\multicolumn{1}{l}{$\sum{}$} & 
\multicolumn{1}{c}{30\%} &
\multicolumn{1}{c}{40\%} &
\multicolumn{1}{c}{30\%} &
 100.00\% \\
\end{tabular}
\end{footnotesize} \end{center}
\end{table}

\begin{figure}[!htbp]
\begin{center} \begin{tabular}{c}
\includegraphics[width=0.7\linewidth]{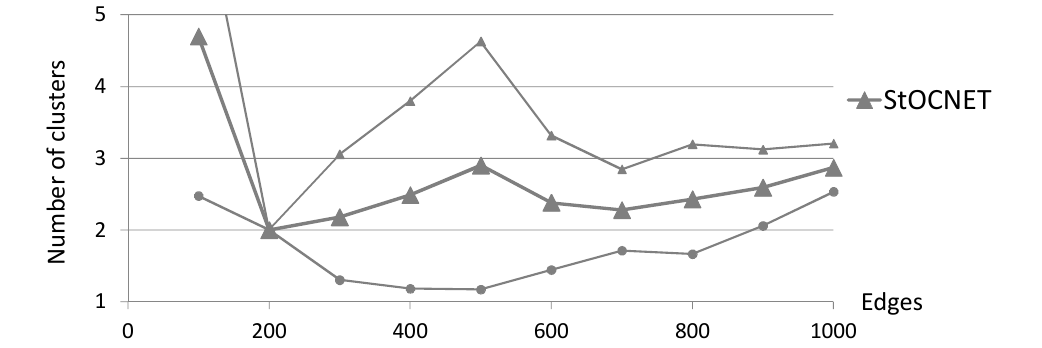} \\
\includegraphics[width=0.7\linewidth]{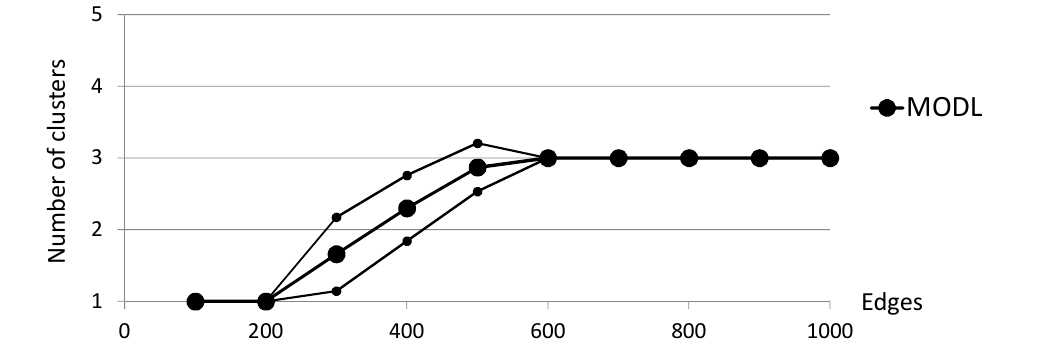} 
	\end{tabular} \end{center}
	\caption{Blockmodel graph: mean number of clusters per sample size, plus/minus the standard deviation}
	\label{figArtificialBM3}
\end{figure}

In Figure~\ref{figArtificialBM3}, we report the mean plus/minus the standard deviation of the selected block number, both for the StOCNET sofware and our approach.
While the clarity criterion of StOCNET selects on average the correct number of clusters for sufficiently large sample size, the selection approach is not resilient to sample variability.
Our approach builds one single cluster  for less than 200 edges and exactly three clusters beyond 600 edges, with a transition between 300 and 500 edges where two clusters (A versus B $\cup$ C) are sometimes selected.

This first experiment shows that our method behaves as a blockmodeling approach w.r.t. the retrieved patterns. 
Being non-parametric, it benefits from a regularized criterion to reliably estimate the correct granularity of the blockmodel pattern. 

\subsubsection{Random Graphs}

The Erd\H{o}s-R{\'e}nyi \cite{ErdosEtAl76} random graph datasets used for tests were introduced by Johnson {\it et al.} \cite{JohnsonEtAl89}.
These graphs are undirected simples graphs, that we treat as symmetric directed graphs with twice the number of edges in the adjacency matrix.
The 16 available random graphs have vertex numbers 124, 250, 500 and 1,000, with average total degree 2.5, 5, 10 and 20.

As StOCNET is limited to graphs with at most 200 vertices, we could apply it only on the random graphs with 124 vertices. 
StOCNET takes 380 seconds on average on each 124 vertices graphs, while our method takes 0.4 second on average for these small graphs and up to 18 seconds for the largest graph with 1,000 vertices and 20,000 edges.

On the 124 vertices graphs, for each average vertex degree, the clarity criterion of StOCNET significantly decreases with the number of blocks, leading to the choice of maximum number of blocks considered (8 in the software). Using this criterion does not lead to a reliable choice of the block number in case of noisy data.
Conversely, on all the random graphs up to 1,000 vertices and 20,000 edges, our method builds one single cluster, which confirms its high resilience to noise.

\medskip
Overall, our method avoids the problem of parametric approaches where the number $K$ of clusters is a user parameter: it neither suffers from under-fitting ($K$ too small) or over-fitting ($K$ too large).

\subsection{Comparison with MDL Coclustering}
\label{secMDLCoclustering}

In this section, we compare our approach with the \emph{Cross-Association} method \cite{ChakrabartiEtAl04}.
As pointed out in Section~\ref{secRelatedWork}, this method is close to our approach: it performs a coclustering on (binary) matrices, is parameter-less and based on MDL, and is scalable, allowing experiments on large datasets.

\subsubsection{Block-Diagonal Graphs}

\begin{figure}[!htbp]
\begin{center}
\includegraphics[width=0.25\linewidth]{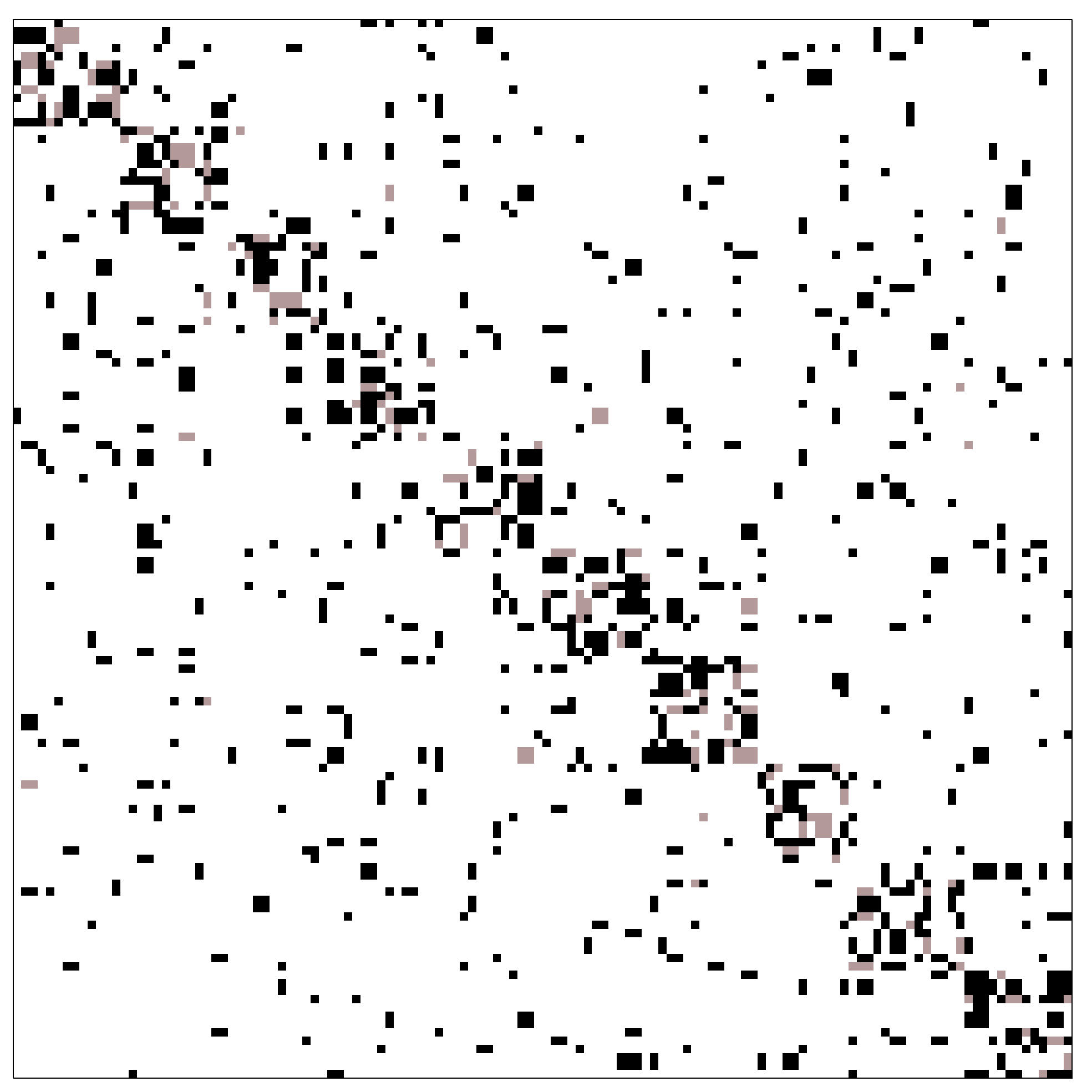} ~~
\includegraphics[width=0.25\linewidth]{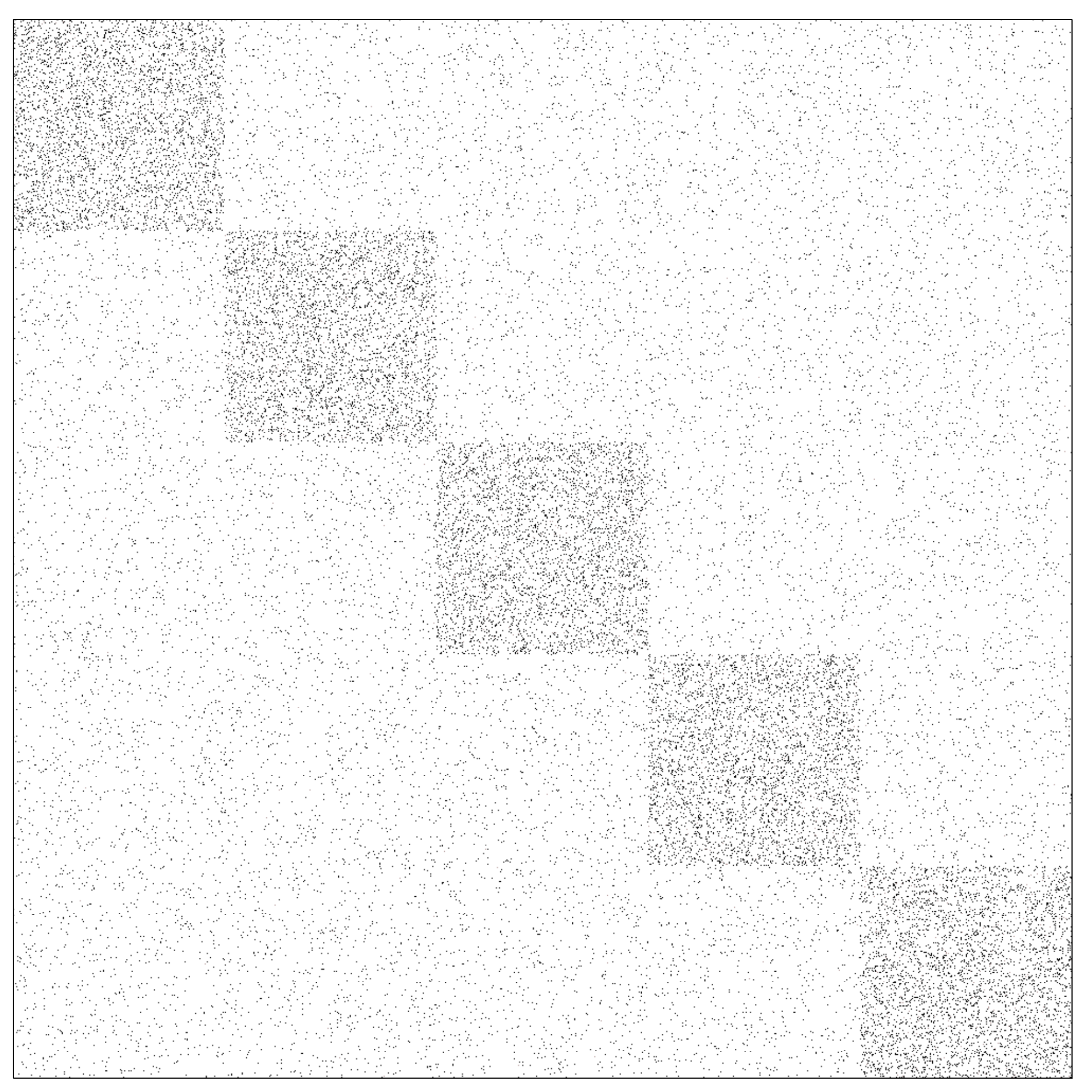} ~~
\includegraphics[width=0.25\linewidth]{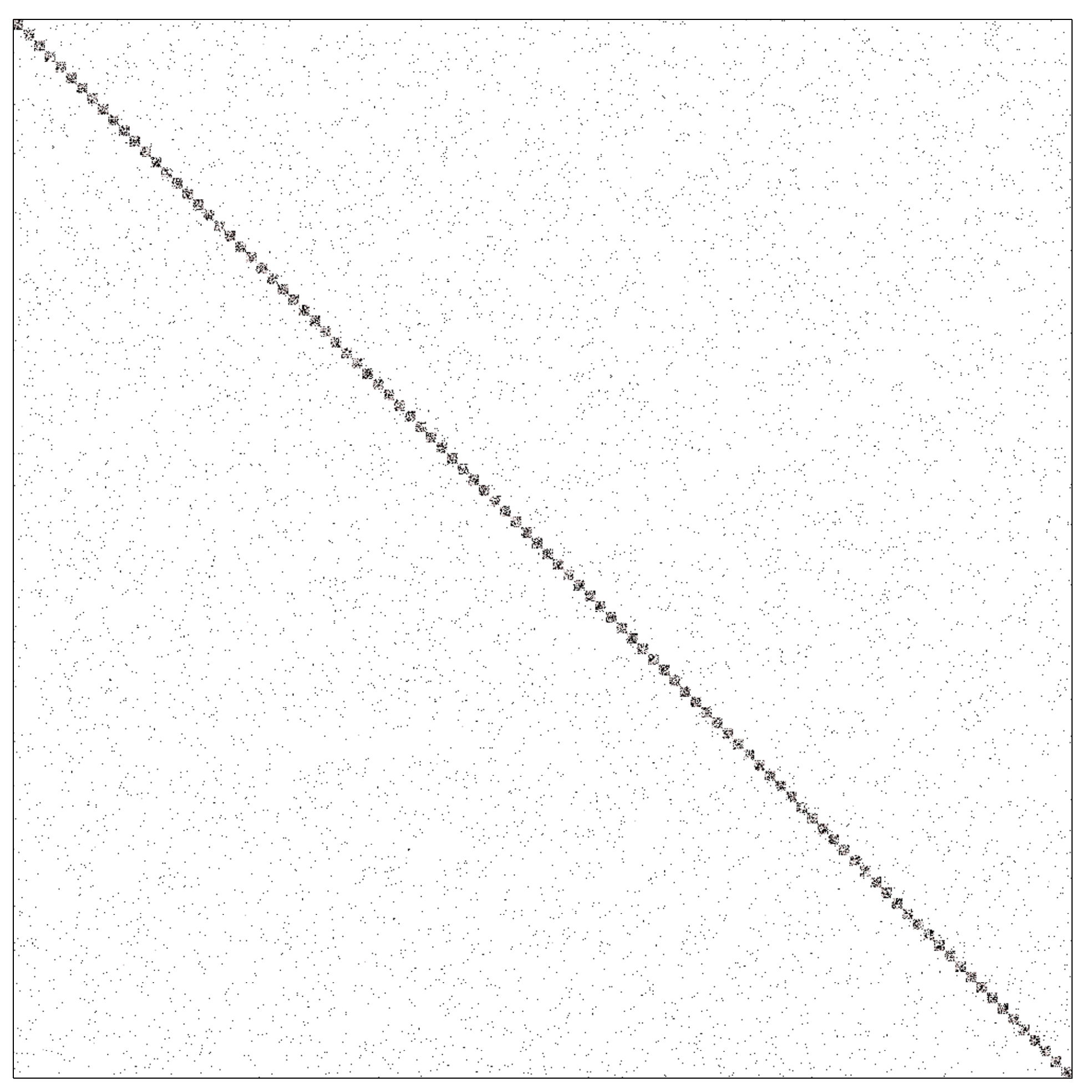}
\end{center}
	\caption{Sample of Noisy(100, 10), Noisy(1000, 5) and Noisy(1000, 100) block-diagonal graphs}
	\label{figArtificialBDGraphSamples}
\end{figure}

Let us introduce \emph{block-diagonal graphs} as directed multigraphs with a partition of the source (resp. target) vertices into $K$ clusters (chosen of equal size in this experiment), with edges lying in the related diagonal blocks. 
For each randomly chosen source vertex, an edge is generated randomly in the target cluster related to same block.
In a noisy version of this pattern, edges are generated randomly among the whole graph with probability $p$, and according to the block-diagonal pattern with probability $(1-p)$. The contingency matrix of some samples of noisy block-diagonal graphs is drawn in Figure~\ref{figArtificialBDGraphSamples}.

In the experiments, we use the same number of source and target vertices and consider the block-diagonal graphs described in Table~\ref{tab:blockdiagonalgraphs}. 
The \emph{Pure} and \emph{Noisy} graphs are block-diagonal graphs with a variety of number of vertices and blocks, whereas the \emph{Random} graphs reduce to Erd\H{o}s-R{\'e}nyi random directed multigraphs.
For all power of 2 sizes 1, 2, 4, 8, ... up to one million edges, we generate ten random graph samples according to the edge distribution of each artificial graph  summarized in Table~\ref{tab:blockdiagonalgraphs}.
We run both the Cross-Association software
\footnote{
The Cross-Association software can be downloaded at
\url{http://www.cs.cmu.edu\-/$\sim$deepay/mywww/software/CrossAssociations-01-27-2005.tgz}.
I am grateful to D.~Chakrabarti for making his method available and providing guidance in its use.
}
and our method with their default settings, and collect the mean cluster number and computation time.

\begin{table}[!htb]
\begin{center}
\begin{footnotesize}
\caption{Block-diagonal graphs}
\begin{tabular}{lccc}
\hline
Name & \#Vertices & \#Blocks & Noise rate \\
\hline
Pure(10, 2) & 10 & 2 & 0\% \\
Pure(100, 10) & 100 & 10 & 0\% \\
Pure(1000, 5) & 1000 & 5 & 0\% \\
Pure(1000, 100) & 1000 & 100 & 0\% \\
Pure(10000, 200) & 10000 & 200 & 0\% \\
Noisy(10, 2) & 10 & 2 & 50\% \\
Noisy(100, 10) & 100 & 10 & 50\% \\
Noisy(1000, 5) & 1000 & 5 & 50\% \\
Noisy(1000, 100) & 1000 & 100 & 50\% \\
Noisy(10000, 200) & 10000 & 200 & 50\% \\
Random(10) & 10 & 1 & 100\% \\
Random(100) & 100 & 1 & 100\% \\
Random(1000) & 1000 & 1 & 100\% \\
Random(10000) & 10000 & 1 & 100\% \\
\hline
\end{tabular}
\end{footnotesize}
\end{center}
\label{tab:blockdiagonalgraphs}
\end{table}

\subsubsection{Cluster Number}

\begin{figure}[!htbp]
\begin{center}
\includegraphics[viewport=90 50 1310 510, width=0.95\linewidth]{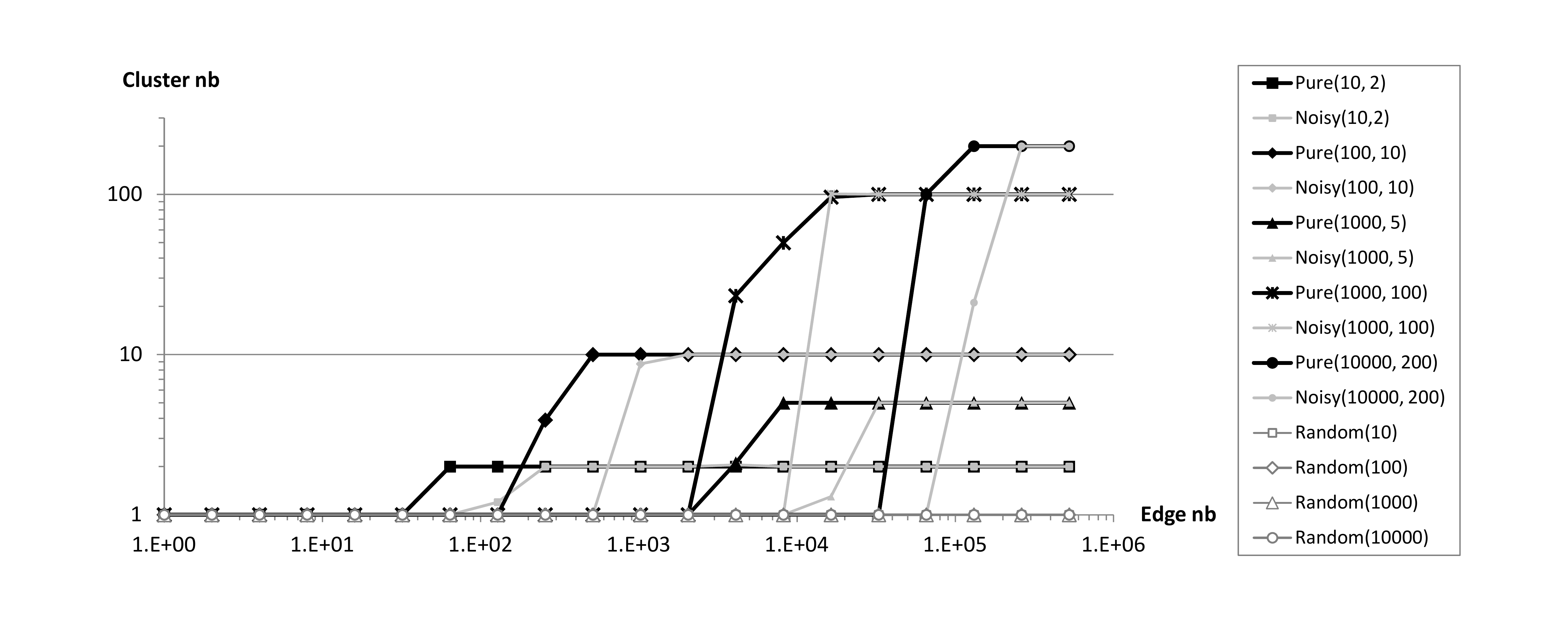} \\
\end{center}
	\caption{Block-diagonal graphs: mean number of clusters per sample size using the MODL method}
	\label{figArtificialBDGMODLCluster}
\end{figure}

\begin{figure}[!htbp]
\begin{center}
\includegraphics[viewport=90 50 1310 510, width=0.95\linewidth]{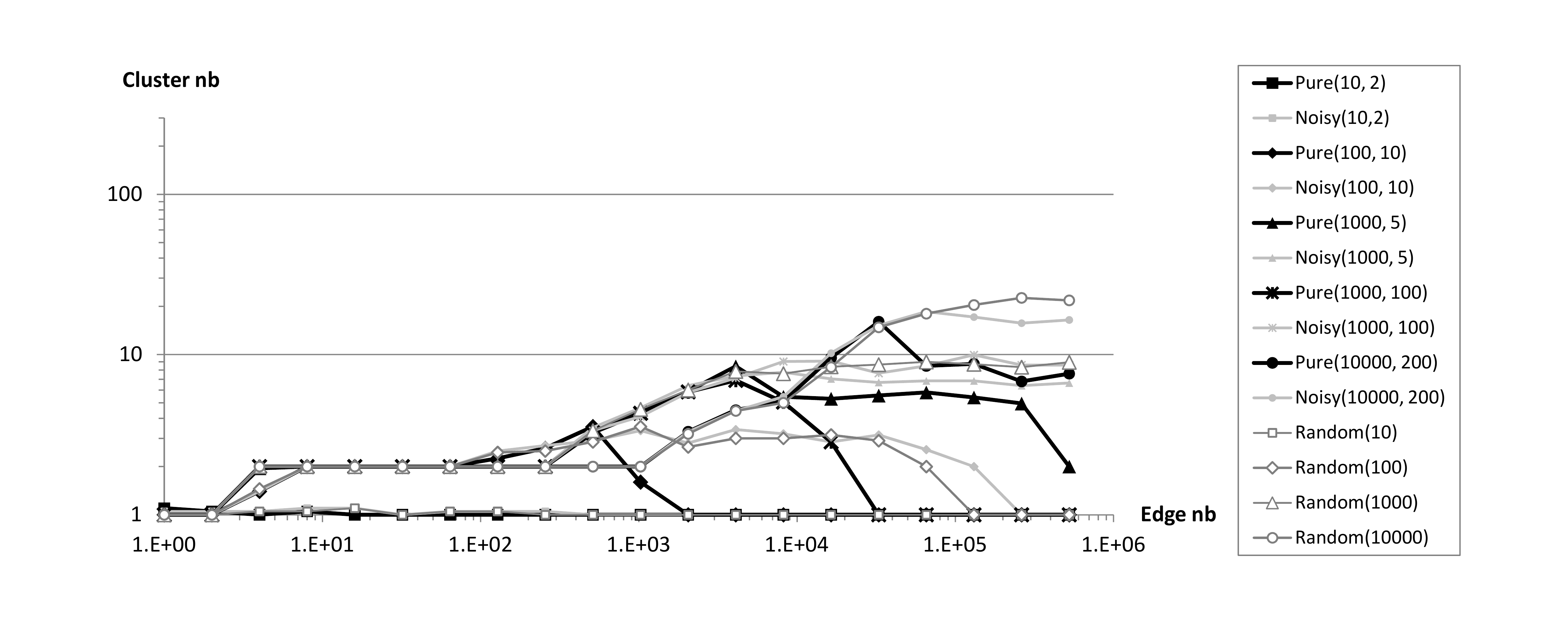} \\
\end{center}
	\caption{Block-diagonal graphs: mean number of clusters per sample size using the Cross-Association method}
	\label{figArtificialBDGCACluster}
\end{figure}

In Figure~\ref{figArtificialBDGMODLCluster} and \ref{figArtificialBDGCACluster}, we report the mean cluster number per sample size obtained with the MODL and Cross-Association methods for each artificial graph summarized in Table~\ref{tab:blockdiagonalgraphs}. 

For each graph, we observe three phases in Figure~\ref{figArtificialBDGMODLCluster} for the MODL method, as in the experimental convergence study in Section~\ref{secConvergenceRate}:
\emph{stability phase}, with one single cluster, \emph{non-parametric estimation phase}, with a growing number of clusters, and \emph{classical estimation phase}, where the true cluster number is retrieved.
The second phase is very sharp for the block-diagonal graphs: below a threshold, which depends on the complexity of the pattern (number of vertices and of blocks), only one cluster is retrieved, and above about four times this threshold, the correct number of clusters is retrieved. The recognition threshold mainly increases with the number of vertices: simpler patterns require less edges to be identified, from 50 edges for the simplest Pure(10, 2) graph to 250,000 edges for the most complex Noisy(10000, 200) graph.
In case of noisy data (with 50\% noise), the shape of the curves is approximately the same, with a translation towards larger sample sizes: the noisy patterns require around four times the edge number necessary to retrieve the related pure patterns.
Finally, the MODL method is highly resilient to noise: it never produces more cluster than expected, and always outputs one single cluster in the extreme case of random graphs.

The results obtained by the Cross-Association method in Figure~\ref{figArtificialBDGCACluster} are quite unclear.
Still, this blurred behavior may be explained by the following reasons.
First, in case of pure patterns, the top-down algorithm can get stuck in local minima, resulting in an under-fitting behaviour in case of patterns with many clusters.
Second, in case of random patterns, as acknowledged in \cite{ChakrabartiEtAl04}, the method tends to pick up spurious patterns and more generally to over-fit the data.
Last, the Cross-Association is designed for binary matrices related to directed simple graphs, not to directed multigraphs. Therefore, as any noisy multigraph is asymptotically flattened to a complete simple graph, the Cross-Association  method produces one single cluster for noisy patterns with many edges.

\subsubsection{Computation Time}

\begin{figure}[!htbp]
\begin{center}
\includegraphics[viewport=90 50 1310 510, width=0.95\linewidth]{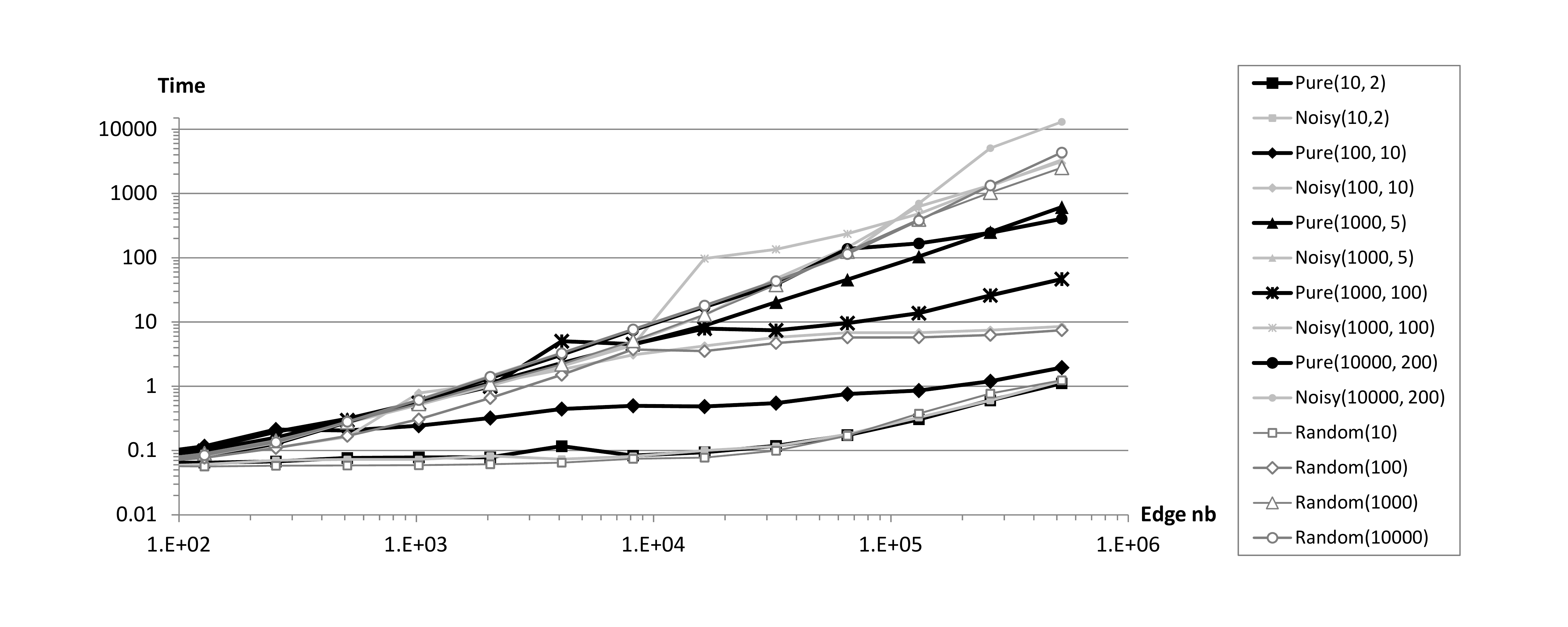} \\
\end{center}
	\caption{Block-diagonal graphs: mean computation time per sample size using the MODL method}
	\label{figArtificialBDGMODLTime}
\end{figure}

\begin{figure}[!htbp]
\begin{center}
\includegraphics[viewport=90 50 1310 510, width=0.95\linewidth]{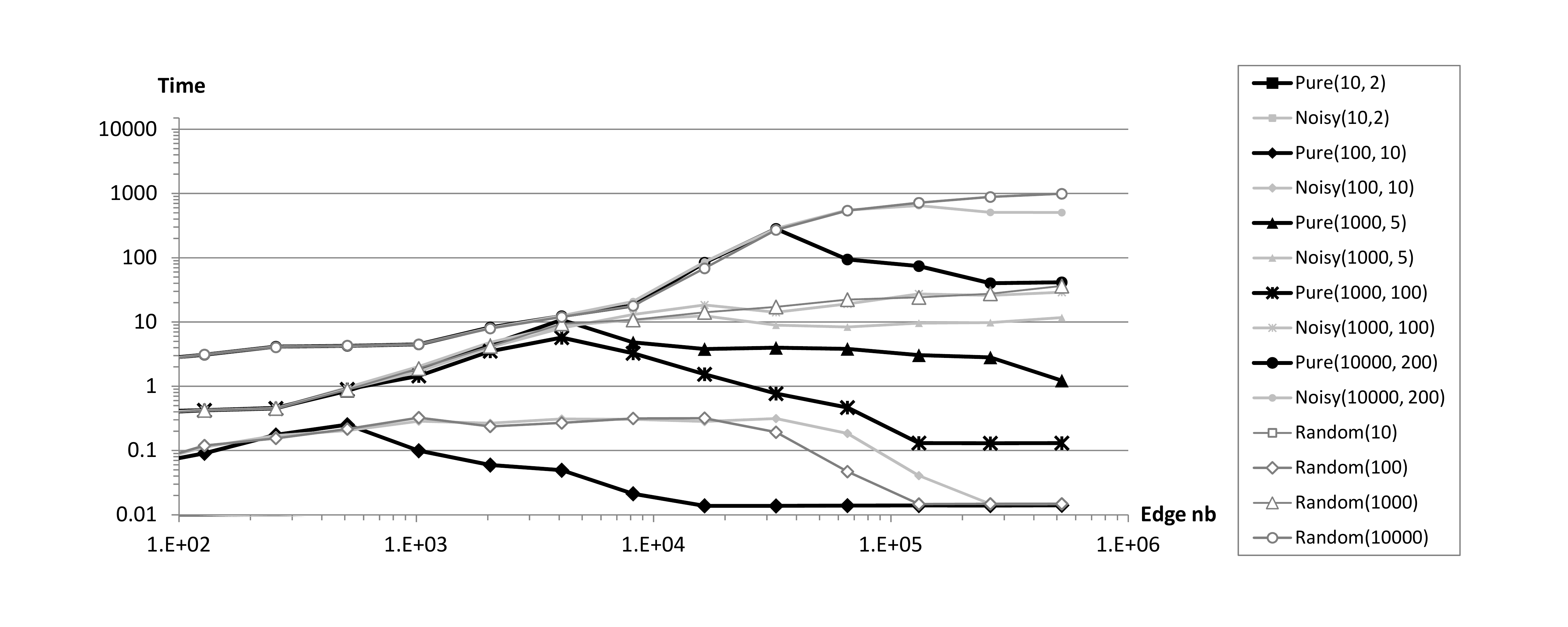} \\
\end{center}
	\caption{Block-diagonal graphs: mean computation time per sample size using the Cross-Association method}
	\label{figArtificialBDGCATime}
\end{figure}

The algorithmic complexity of the MODL method (main greedy bottom-up heuristic described in Algorithm~\ref{GBUM}) is $\mathcal{O}(m \sqrt m \ log m)$, where $m$ is the number of edges.
According to \cite{ChakrabartiEtAl04}, the algorithmic complexity of the Cross-Association method is $\mathcal{O}(m (k^*+l^*)^2)$, where $k^*$ and $l^*$ are the number of source and target clusters retrieved by the top-down heuristic. Although this is quadratic w.r.t. the cluster number, this may be more efficient that our bottom-up approach in case of patterns with small number of clusters.
In Figure~\ref{figArtificialBDGMODLTime} and \ref{figArtificialBDGCATime}, we report the mean computation time per sample size obtained with the MODL and Cross-Association methods. 

Compared to the theoretical computation complexity, the actual computation time depends on many factors, such as the numbers of vertices, the size of the true pattern and the level of noise. 
In the case of the MODL method (Figure~\ref{figArtificialBDGMODLTime}), the shape of the computation time curve in the worse case is compliant with the theoretical complexity, with up to ten thousand seconds for the largest noisy graph.
In the case of the Cross-Association method (Figure~\ref{figArtificialBDGCATime}), the shape of the curves grows quadratically as expected, but is flattened in the end with no more than one thousand seconds for the largest noisy graphs.
This behavior comes from the bottom-up heuristic which is stuck in local minima and hardly produces more than around 20 clusters, which allows the method to be quicker at the expense of under-fitting the data.

\medskip
Overall, whereas our method and the Cross-Association methods are both scalable, parameter-less and based on a similar MDL-based model selection approach, the modeling choices are quite different, resulting in contrasting behavior.
Our method is valid both non-asymptotically and asymptotically and neither suffers from under-fitting nor over-fitting.
It never produces spurious clusters in case of random graphs and recovers complex patterns with an increased precision as the amount of available data increases.

\section{Experiments on Real World Datasets}
\label{secRealWorldExperiments}

This section compares our method with the Cross-Association method \cite{ChakrabartiEtAl04} on real world datasets coming from three different domains: document clustering, webspam detection and exploratory analysis of a flight trip dataset.

\subsection{Document Dataset}
\label{secClassic}

We exploit the CLASSIC3 document dataset
\footnote{
Dataset available at \url{http://www.dataminingresearch.com/index.php/2010/09/classic3-classic4-datasets/}}
introduced in \cite{DhillonEtAl03} to evaluate the Information-Theoretic Coclustering \emph{ITC} method.
This collection of documents comes from three domains: MEDLINE consists of 1,033 abstracts from
medical journals, CISI consists of 1,460 abstracts from information retrieval papers and CRANFIELD consists of 1,398 abstracts from aerodynamic systems.
Overall, this dataset can be represented as a directed multigraph with 3,891 source vertices (documents), 5657 target vertices (words) and 287,827 edges (184,772 edges in a flattened binary representation of the graph).

We build a coclustering of the dataset using the MODL method described in Section~\ref{secMODLforGraphs}.
The running time is one hour and 50 minutes.
We obtained a very fine-grained summary of the dataset, with 151 clusters of documents and 293 clusters of word.

\paragraph*{\bf{Agglomerative hierarchical clustering method}}
In order to explore the coclustering at different granularities, we suggest to coarsen the obtained coclustering by the mean of an agglomerative hierarchical clustering.
We use the following similarity between two clusters $i$ and $j$
\begin{equation}
\label{ClusterSimilarity2}
\begin{split}
\delta (i, j) = & c(M_{i \cup j}) - c(M),\\
  = & \log \frac {p(M | G)} {p(M_{i \cup j} | G)}.
\end{split}
\end{equation}
where $c(M)$ is the evaluation criterion (\ref{E:EDC}) introduced in Section~\ref{secMODLforGraphs}, $M$ is the current coclustering model and $M_{i \cup j}$ is the coclustering model after the merge of clusters $i$ and $j$.
Intuitively, if two clusters are similar, the total code length of the data (see criterion $c(M)$) is not much different between the cases where the clusters are coded jointly or separately, so that $\delta$ will be small.
Actually, the agglomerative hierarchical clustering algorithm is the same as that of Algorithm~\ref{GBUM}, except that the merges are performed until the required number of clusters is obtained.

\begin{figure}[!htbp]
\begin{center} 
	\includegraphics[width=0.45\linewidth]{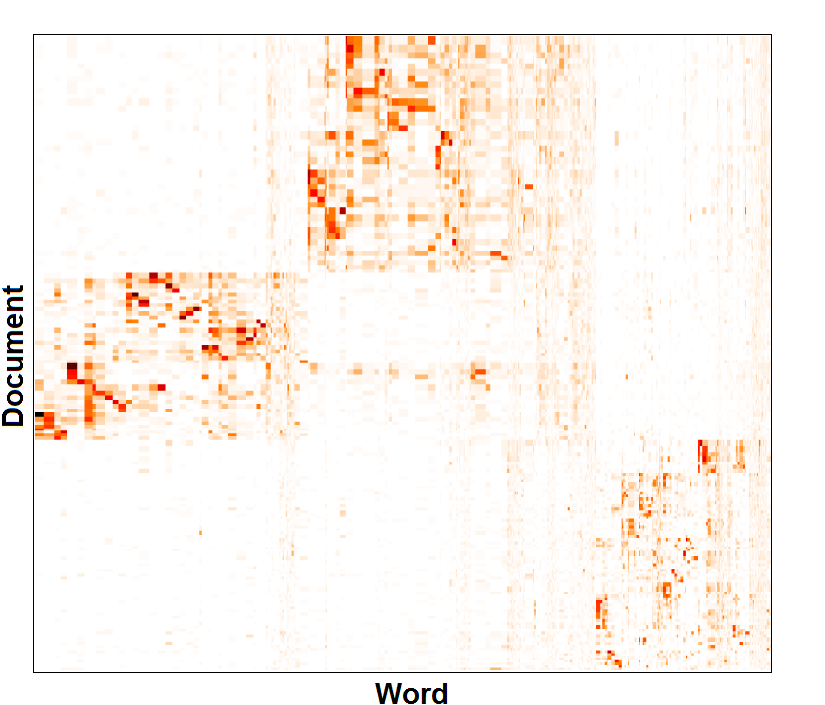}
	\includegraphics[width=0.45\linewidth]{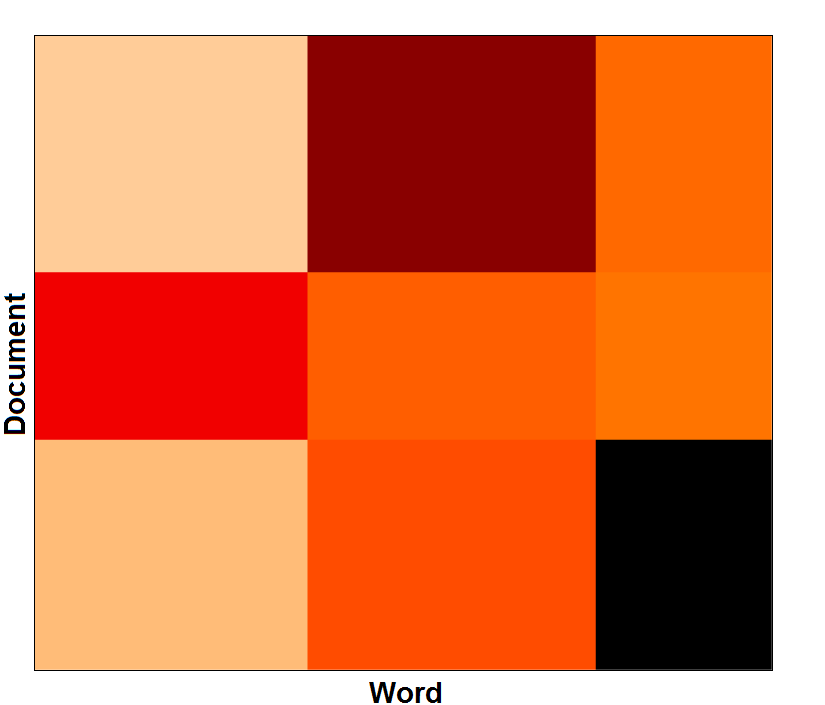}
\end{center}
	\caption{Coclustering of the CLASSIC3 dataset: initial fine-grained $151*293$ coclustering matrix and coarse $3*3$ matrix.}
	\label{classic3Coclustering}
\end{figure}

\medskip
As the CLASSIC3 dataset comes from three domains, we chose to coarsen our fine-grained $151*293$ coclustering matrix down to a 3*3 coarse matrix. The fine and coarse grained matrix are shown in Figure~\ref{classic3Coclustering}.

Applying the Cross-Association \emph{CA} method, we obtain a 20*20 summary of the flattened binary dataset in about six minutes.

We evaluate the agreement between the coclustering results and the known documents classes for the MODL and CA methods:
\begin{itemize}
	\item MODL: we exploit the coarse 3*3 matrix to correlate the three obtained clusters of document with the known documents classes,
	\item CA: we collect the majority class in each obtained cluster of document and manually group the clusters sharing the same majority class, down to three clusters.
\end{itemize}

The results are reported in Figure~\ref{figClassic3Agreement}.
In \cite{DhillonEtAl03}, the ITC method is applied with three clusters as a user parameter, and obtains a good agreement between the retrieved clusters and the true classes, with about 60 agreement errors.
The CA methods obtain a better agreement than the ITC method, with about $20\%$ less errors.
The results are better for the MODL method with twice less agreement errors than for the CA method.

\begin{figure}[!htb]
\begin{center}
\hspace{-1.2cm}
\begin{minipage}[c]{.4\linewidth} \begin{center} \begin{scriptsize} \begin{tabular}
{@{}r|c|c|c|@{}l} 
\multicolumn{1}{c}{\textbf{MODL}} & \multicolumn{1}{c}{MED} & \multicolumn{1}{c}{CISI} & \multicolumn{1}{c}{CRAN} & $\sum{}$ \\ \cline{2-4}
C1 & 1025	& 2	   & 0	  & 1027 \\ \cline{2-4}
C2 & 5	  & 1446 & 0	  & 1451 \\ \cline{2-4}
C3 & 3	  & 12	 & 1398	& 1413 \\ \cline{2-4}
\multicolumn{1}{r}{$\sum{}$} & \multicolumn{1}{c}{1033} & \multicolumn{1}{c}{1460} & \multicolumn{1}{c}{1398} & 3891 \\ 
\end{tabular} \end{scriptsize} \end{center} \end{minipage}
\begin{minipage}[c]{.12\linewidth} 
~
 \end{minipage}
\begin{minipage}[c]{.4\linewidth} \begin{center} \begin{scriptsize} \begin{tabular}
{@{}r|c|c|c|@{}l} 
\multicolumn{1}{c}{\textbf{CA}} & \multicolumn{1}{c}{MED} & \multicolumn{1}{c}{CISI} & \multicolumn{1}{c}{CRAN} & $\sum{}$ \\ \cline{2-4}
C1 & 1014	& 9	   & 3	  & 1026 \\ \cline{2-4}
C2 & 17	  & 1450 & 16	  & 1483 \\ \cline{2-4}
C3 & 2	  & 1	   & 1379	& 1382 \\ \cline{2-4}
\multicolumn{1}{r}{$\sum{}$} & \multicolumn{1}{c}{1033} & \multicolumn{1}{c}{1460} & \multicolumn{1}{c}{1398} & 3891 \\ 
\end{tabular} \end{scriptsize} \end{center} \end{minipage}
\end{center}
	\caption{CLASSIC3: agreement with the true classes for the MODL and CA methods.}
	\label{figClassic3Agreement}
\end{figure} 

Overall, the MODL method produces a finer coclustering than the CA method, at the expense of a a higher computation time. This finer model better fits the data, with a better agreement with the true classes in the dataset.

\subsection{Web Spam Dataset}
\label{secChallenge}

In this section, we evaluate the benefit of our method as a preprocessing step for web spam detection.

\subsubsection{Web Spam Challenge 2007}

\begin{figure}[!htbp]
\begin{center} 
	\includegraphics[viewport=70 70 700 350, width=0.70\linewidth]{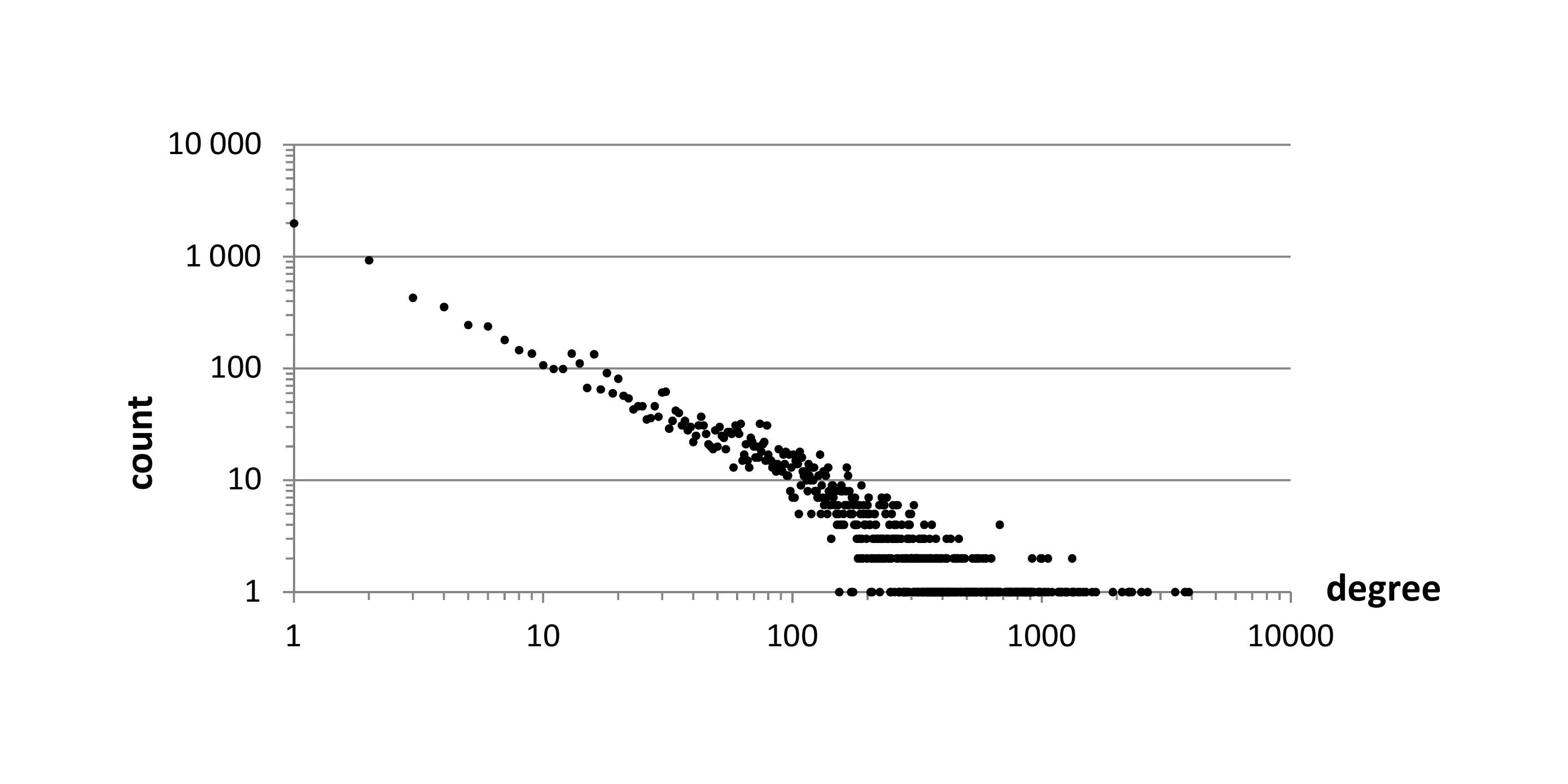}
\end{center}
	\caption{Webspam challenge host graph: number of vertices per output degree.}
	\label{webspamVertexOutDegrees}
\end{figure}

Web spam consists in manipulating the relevance of resources indexed in a manner inconsistent with the purpose of the indexing system of internet search providers.
The data used in this paper comes from the Web Spam Challenge (corpus \#1, Track II) \footnote{\url{http://webspam.lip6.fr/}, Web Spam Challenge 2007} held in conjunction with the 2007 ECML/PKDD Graph Labeling workshop.
The goal of the challenge is to evaluate machine learning methods to label web hosts to be spam or normal.
The data consists of 9,072 web hosts with both content and link data.
Content data correspond to the TF-IDF vectors over 100 web pages of the host, with almost 5 millions features. 
The hosts are the vertices and the link data represent the edges in the directed host graph, with one edge per hyperlink between two hosts. Overall, the host graph contains 514,700 edges, all of them are simple edges. The degrees of the vertices follow the power-law distribution, as shown in Figure~\ref{webspamVertexOutDegrees}.
The training set consists of 907 hosts labelled as normal or spam, with approximately $20\%$ spam.
The objective of the challenge is to label the test set (8,165 hosts). The result is assessed using the area under the ROC curve (AUC).

The results of the challenge participants.
\footnote{
Available at \url{http://airweb.cse.lehigh.edu/2008/web\_spam\_challenge/introduction.pdf}.
Computation time of the participants is not available.
}
are reported on the right of Figure~\ref{webspamResultAnalysis}.
They all used both the content and link data in a semi-supervised learning setting, on top of classifiers among which SVM, random forests and naive Bayes.
All the available data is exploited to build a model: train and test content and link data, plus the train labels.
The link data is exploited either by extracting link-based features, such as number (or ratio) of links from (or to) to spam or normal hosts, or by constraining the classifier to account for the labels of the connected hosts.

\subsubsection{Evaluation}

\begin{figure}[!htbp]
\begin{center} 
	\includegraphics[width=0.60\linewidth]{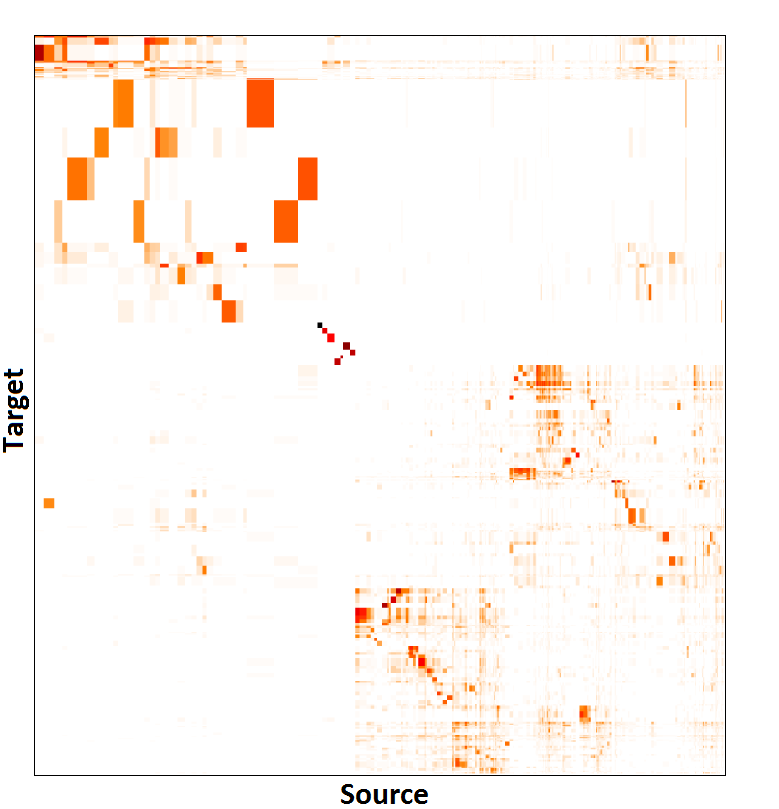}
\end{center}
	\caption{Coclustering of the host graph of the web spam challenge.}
	\label{webspamGraph}
\end{figure}

In a first step, we exploit the link data only, that is the directed graph consisting of 9,072 hosts with 514,700 links.
All the hosts and links are processed without any output label, to identify the ``natural'' clusters of source and target hosts.
We build a coclustering of the source and target hosts using the MODL method described in Section~\ref{secMODLforGraphs}.
The running time is 3 hours and 14 minutes.
The coclustering of the host graphs retrieves 167 clusters of source hosts and 219 clusters of target hosts.
The coclustered matrix is displayed in Figure~\ref{webspamGraph}, which shows that the graph of hosts is highly structured, with most of the information lying in few hundred of coclusters.
The asymmetry in the hyperlinks of the host graph conforms to the observation of the challenge participants, that there is usually no link from a normal host to a spam host.

Applying the alternative Cross-Association method, we obtain a 10*10 coclustering of the dataset in about five minutes.

\begin{figure}[!htbp]
\begin{center} 
	\includegraphics[viewport=140 70 440 210, width=0.9\linewidth]{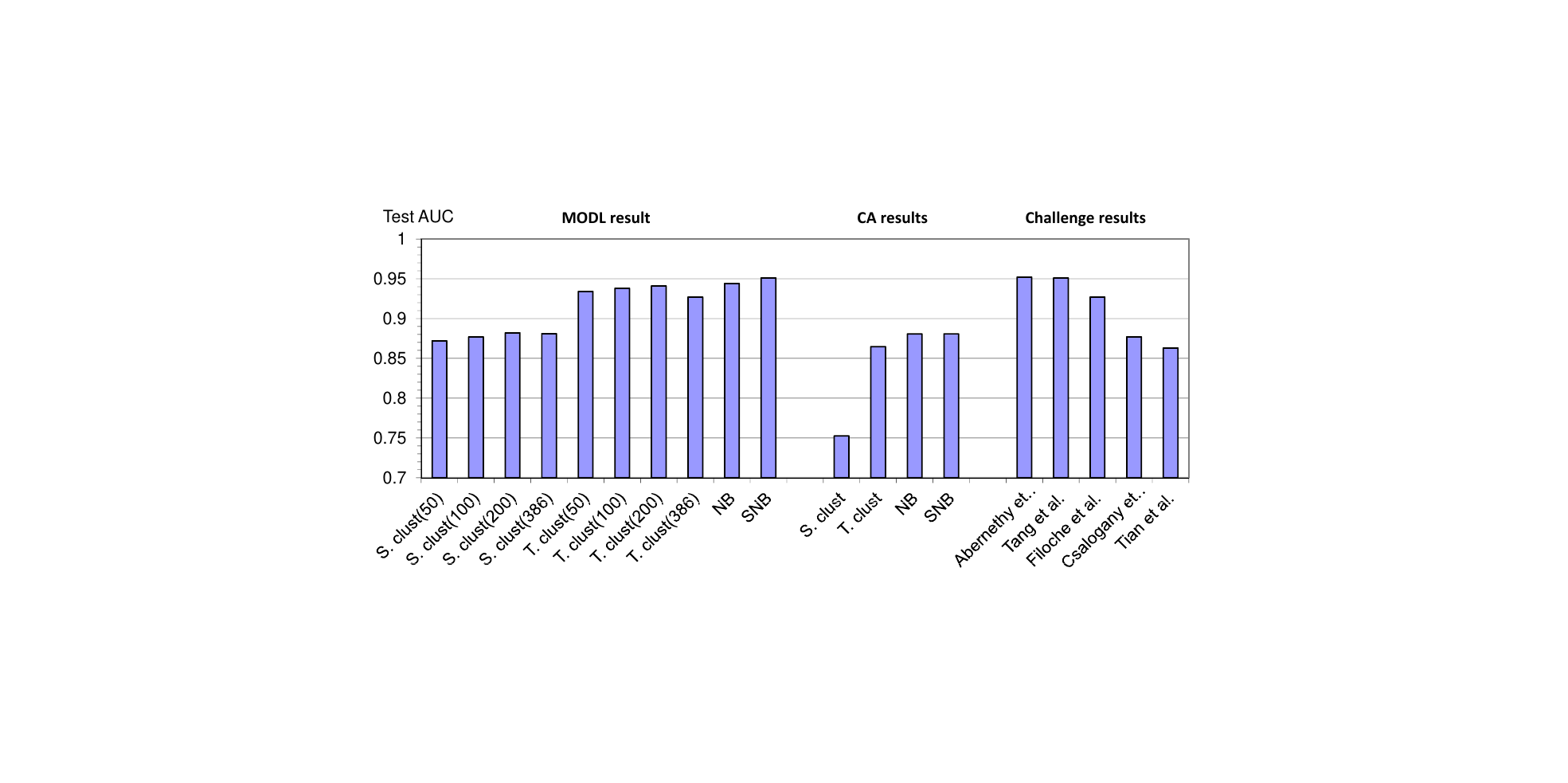}
\end{center}
	\caption{Webspam 2007, phase II: test AUC results for the MODL and CA methods, and of the challenge participants.}
	\label{webspamResultAnalysis}
\end{figure}

In a second step, the available labeled train hosts are used to describe the output distribution in each cluster of hosts.
The label of a test host is then predicted according to the output distribution of its cluster.
However, the MODL coclustering is fine grained, and about $20\%$ of the clusters do not contain a single train instance: in this case the test host is labeled as the majority class (normal).
To alleviate this problem, we chose to coarsen our coclustering at different grain levels with the agglomerative hierarchical clustering method described in Section~\ref{secClassic}.
We build three new coarsened coclusterings, starting from the initial $386=167+219$ clusters down to 200, 100 and 50 clusters.
Given this preprocessing, each host is represented by eight variables, source or target cluster at four grain levels.
On the left of Figure~\ref{webspamResultAnalysis}, we report the test AUC obtained by each univariate cluster-based classifier for the MODL coclustering (\emph{S. clust($k$)} and \emph{T. clust($k$)} for classifiers based on source or target clusters for coclusterings with $k$ clusters). We also combined these classifiers using a Naive Bayes (NB) and a Selective Naive Bayes (SNB) classifier \cite{BoulleJMLR07}.
Using the same protocol, we report the test AUC results obtained using the $10$ times $10$ coclustering retrieved by the Cross-Association method on the center of Figure~\ref{webspamResultAnalysis}.
Finally, we also report the results of the challenge participants on the right of Figure~\ref{webspamResultAnalysis}.
The test AUC results are obtained using the predefined train/test split of the dataset to allow a comparison 
with the challenge participants.
No cross-validation results are available for the challenge participants.
Given that the size of the test set is $s=8165$, the expected variance of the results is around $\frac {1} {\sqrt s} \approx 1\%$.

The Cross-Association method under-fits the data, with an insufficient number of clusters to correctly identify the spam class.
For both the MODL and CA methods, the results show that the target clusters are much more informative than the source clusters, with $5$ to $10\%$ better test AUC.
The granularity of the coclustering has a medium impact on the MODL test AUC: the best results is obtained with about 200 clusters.
The combined classifiers manage to exploit all the input variables to get superior performance.
It is noteworthy that the (MODL) SNB classifier obtains a test AUC comparable to that of the winner of the challenge, while using the link data only.

\smallskip
Like in the document clustering experiment, the MODL method produces a finer coclustering than the CA method, at the expense of a a higher computation time. This finer model better fits the data, with competitive performance for the task of web spam prediction.

\subsection{Flight Trip Dataset}
\label{secAirline}

In this section, we apply our method on a large passenger flight trip dataset for the purpose of exploratory analysis, when no ground truth is available.
The dataset
\footnote{\url{http://www.transtats.bts.gov/DataIndex.asp}, Airline Origin and Destination Survey (DB1B) 2010,
 RITA: Bureau of Transportation Statistics}
comes from the US Bureau of Transportation Statistics and contains a 10\% sample of all airline tickets for each domestic flight trip.
We collected the origin and destination fields as well as the number of passengers per trip for the four terms of year 2010, so as obtain 10\% of all US domestic passenger flight trips for one full year.
The resulting data table contains 22,038,685 records, for a total of 469 airports and 43,092,916 flight trips. 
The airports are the vertices and the trips are the oriented multiple edges from origin to destination airports.
The average edge multiplicity in this graph is 471, with 91,482 different edges.

We apply the MODL coclustering method on this dataset, with a running time of 16 minutes 40 seconds.
We obtain a quasi symmetric summary of the multigraph with 233 clusters of origin airports, 234 clusters of destination airports, and 46,325 non empty coclusters.

\begin{table}[!htb]
\label{tableUSAirTripsCoclustering}
\begin{center} \begin{footnotesize} 
\caption{US flight trips and their coclustering summary.}
\begin{tabular}{l|c|c|c|c|c|c} 
\multicolumn{1}{l}{} & 
\multicolumn{1}{c}{PI} &
\multicolumn{1}{c}{WC} &
\multicolumn{1}{c}{FL} &
\multicolumn{1}{c}{DV} &
\multicolumn{1}{c}{CE} &
 $\sum{}$ \\ \hhline{~|-|-|-|-|-|~}  
PI	& 1.03\%	& 0.92\%	& 0.03\%	& 0.09\%	& 0.49\%	& 2.56\% \\ \hhline{~|-|-|-|-|-|~} 
WC	& 0.89\%	& 17.60\%	& 1.79\%	& 3.64\%	& 10.54\%	& 34.46\% \\ \hhline{~|-|-|-|-|-|~} 
FL	& 0.03\%	& 1.69\%	& 0.30\%	& 2.74\%	& 6.15\%	& 10.91\% \\ \hhline{~|-|-|-|-|-|~} 
DV	& 0.09\%	& 3.77\%	& 3.13\%	& 0.36\%	& 6.41\%	& 13.76\% \\ \hhline{~|-|-|-|-|-|~} 
CE	& 0.28\%	& 11.06\%	& 5.60\%	& 6.31\%	& 15.07\%	& 38.32\% \\ \hhline{~-----~} 
\multicolumn{1}{l}{$\sum{}$} & 
\multicolumn{1}{c}{2.32\%} &
\multicolumn{1}{c}{35.04\%} &
\multicolumn{1}{c}{10.85\%} &
\multicolumn{1}{c}{13.14\%} &
\multicolumn{1}{c}{38.66\%} &
 100.00\% \\
\end{tabular}
\end{footnotesize} \end{center}
\end{table}
 
\begin{figure}[!htbp]
\begin{center} 
	\includegraphics[width=0.97\linewidth]{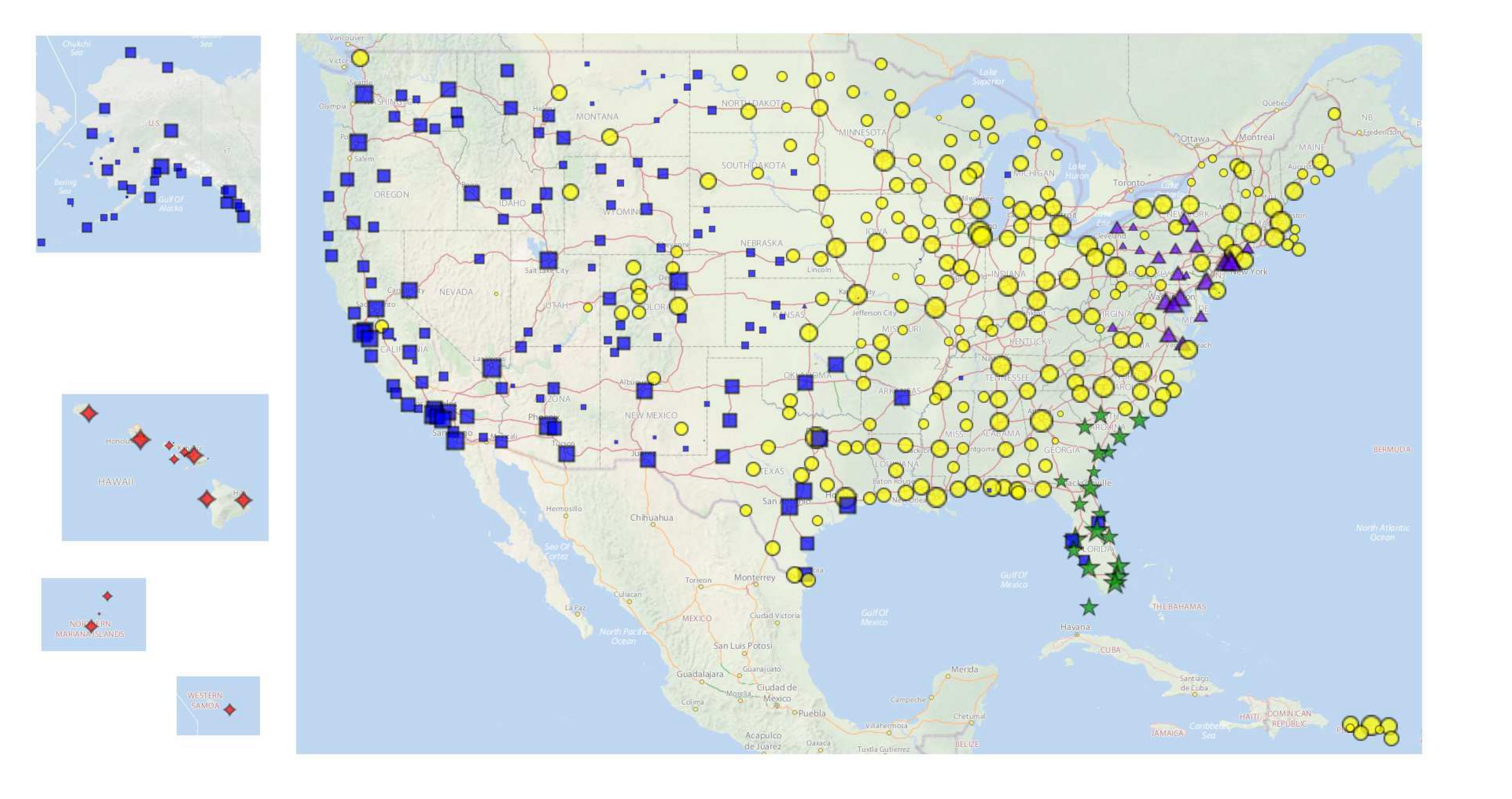}
\end{center}
	\caption{US flight trips: five clusters of airports.}
	\label{USCarrier5Clusters}
\end{figure}

In order to explore the extracted coclustering with a broader picture, we chose to coarsen our coclustering with the agglomerative hierarchical clustering method described in Section~\ref{secClassic}.
For depiction purpose, we keep five clusters: the coclustering matrix summary is almost symmetric, as shown in Table~\ref{tableUSAirTripsCoclustering}.
Figure~\ref{USCarrier5Clusters} displays the five clusters of destination airports. 
There is a clear geographic correlation in the clusters. A first cluster consists of the Pacific Islands (PI), with Hawai, Mariana Islands, Guam and Samoa. 
A second cluster roughly corresponds to the Delaware Valley (DV), with counties from  Pennsylvania, New Jersey, Delaware and Maryland. 
A third cluster contains the Florida airports (FL), while the two last clusters correspond the West Coast (WC) and Center and East (CE) of America.

Interestingly, the Pacific Islands cluster is very dense, with 17 times more intra-cluster flight trips than expected in case of independence of the origin and destination of the trips ($1.03\% \approx 17 * (2.56\% * 2.32\%)$).
On the contrary, the Florida and Delaware Valley clusters are sparse clusters, with respectively four times and five times less intra-cluster trips than expected in case of independence. Although these two sparse clusters are not geographically connected, they are linked by twice the number of trips than expected.
This kind of exploratory analysis can be performed at any granularity up to the finest coclustering retrieved by our method.

\smallskip
This illustrates how the MODL coclustering method can be used for the task of exploratory analysis, when no ground truth is available.

\section{Conclusion}
\label{secConclusion}

In this paper, we have presented a novel way of discovering structures in graphs, by considering graphs as generative models whose statistical units are the edges, with unknown joint density of the source and target vertices.
Our method applies the MODL approach based on data grid models \cite{BoulleHOPR10} to the case of directed multigraphs. By clustering both the source and target vertices of a graph, the method behaves as a non-parametric estimator of the unknown edge density.
The modeling approach exploits the MDL principle in a data-dependent way: it aims to model the finite graph sample directly. The modeling task is then easier, with finite modeling space and model prior which essentially reduce to counting. This modeling approach is both non-asymptotic and consistent, with an asymptotic convergence towards the true edge density, without any assumption regarding this density.

Experiments on artificial data demonstrate that our approach is both robust, building one single cluster in case of random graphs, and accurate, being able to recover fine grained patterns.
Our method has been applied to a document classification problem and to a Web spam detection problem in a graph of hosts. The patterns retrieved by our approach are highly informative, with a good agreement with the true classes.
An application on a large flight trip dataset shows the potential interest of our method for exploratory analysis, with the extraction of insightful and interpretable patterns.

Our method has a $\mathcal{O}(m \sqrt m \log m)$ time complexity, where $m$ is the number of edges, providing practical running times up to millions of edges.
In future work, we plan to work on faster and more scalable optimization algorithms in order to deal with very large graphs, up to billions of edges.

\appendix
\section*{APPENDIX}
\setcounter{section}{1}

\subsection{Proof of theorems}
\label{AppendixA}

In this appendix we prove theorems \ref{EdgeDensityCriterionAsymptotics} and \ref{EdgeDensityCriterionLimit} from Section~\ref{secMODLAsymtoticConsistency}.
\smallskip
 
\noindent
{\bf Theorem~\ref{EdgeDensityCriterionAsymptotics}}
{\it
The MODL evaluation criterion (Equation~\ref{E:EDC}) for a graph coclustering model $M$ is asymptotically equal to $m$ times the entropy of the source and target vertex variables minus the mutual entropy of the variables grouped.

\begin{equation*}
c(M) = m \left( H(V_S) + H(V_T) - I(V_S^M; V_T^M) \right) + \mathcal{O}(\log m).
\end{equation*}
}

\begin{proof}
According to Equation~\ref{E:EDC}, we have:
{\allowdisplaybreaks
\begin{subequations} \label{E:PEDC}
\begin{align}
c(M) = & \log n_S + \log n_T  \label{E:PEDC1} \\ 
&+ \log B(n_S, k_S) + \log B(n_T, k_T)  \label{E:PEDC2} \\
& + \log \binom {m + k_E - 1} {k_E - 1}  \label{E:PEDC3} \\
&  + \sum_{i=1}^{k_S} {\log \binom {m_{i.}^S + n_i^S - 1} {n_i^S - 1}}  \label{E:PEDC4} \\
&  + \sum_{j=1}^{k_T} {\log \binom {m_{.j}^T + n_j^T - 1} {n_j^T - 1}}  \label{E:PEDC5} \\
& + \log m! - \sum_{i = 1}^{k_S} {\sum_{j = 1}^{k_T} {\log m_{ij}^{ST}!}}  \label{E:PEDC6} \\
& + \sum_{i = 1}^{k_S} {\log m_{i.}^S!} - \sum_{i = 1}^{n_S} {\log m_{i.}!}  \label{E:PEDC7} \\
&  + \sum_{j = 1}^{k_T} {\log m_{.j}^T!} - \sum_{j = 1}^{n_T} {\log m_{.j}!}  \label{E:PEDC8} 
\end{align}
\end{subequations}}

We study the asymptotic behavior of the criterion when the number of edges grows to infinity, for a fixed set of vertices.
Lines (\ref{E:PEDC1}) and (\ref{E:PEDC2}) of criterion $c(M)$ are bounded by constants w.r.t. $m$.
Since the numbers of clusters ($k_S, k_T$) and the numbers of vertices per cluster ($n_i^S, n_j^T$) are bounded by the number of vertices ($n_S, n_T$), lines (\ref{E:PEDC3}), (\ref{E:PEDC4}) and (\ref{E:PEDC5}) of the criterion, corresponding to the encoding of the model prior parameters, are bounded by $\mathcal{O}(\log m)$.

We now focus on the likelihood terms of the criterion (lines (\ref{E:PEDC6}), (\ref{E:PEDC7}) and (\ref{E:PEDC8}).
Using the approximation $\log n! = n(\log n - 1) + \mathcal{O}(\log n)$ based on Stirling's formula and rearranging the terms with new $m \log m$ terms, we get:

\begin{equation}
\begin{split}
c(M) = & \left(m \log m - \sum_{i = 1}^{k_S} {\sum_{j = 1}^{k_T} {m_{ij}^{ST} \log m_{ij}^{ST}}} \right) \\ 
& - \left(m \log m - \sum_{i = 1}^{k_S} {m_{i.}^S \log m_{i.}^S}\right) \\ 
&  - \left(m \log m - \sum_{j = 1}^{k_T} {m_{.j}^T \log m_{.j}^T}\right)\\ 
&  + \left(m \log m - \sum_{i = 1}^{n_S} {m_{i.} \log m_{i.}}\right) \\ 
&  + \left(m \log m - \sum_{j = 1}^{n_T} {m_{.j} \log m_{.j}}\right) \\
&  + \mathcal{O}(\log m).
\end{split}
\end{equation}

Given that the sum of the edge counts in each partition (per cocluster, per cluster in and out-degree and per vertex in and out-degree) is always equal to $m$, we obtain:
\begin{equation}
\begin{split}
c(M) = & -m \sum_{i = 1}^{k_S} {\sum_{j = 1}^{k_T} {\frac {m_{ij}^{ST}} {m} \log \frac {m_{ij}^{ST}} {m}}} \\
&  + m \sum_{i = 1}^{k_S} {\frac {m_{i.}^S} {m} \log \frac {m_{i.}^S} {m}} 
  + m \sum_{j = 1}^{k_T} {\frac {m_{.j}^T} {m} \log \frac {m_{.j}^T} {m}}  \\
&  - m \sum_{i = 1}^{n_S} {\frac {m_{i.}} {m} \log \frac {m_{i.}} {m}} 
   - m \sum_{j = 1}^{n_T} {\frac {m_{.j}} {m} \log \frac {m_{.j}} {m}} \\
&  +  \mathcal{O}(\log m).  
\end{split}
\end{equation}

As the marginal distributions $m_{i.}^S$ and $m_{.j}^T$ can be decomposed by summation on the joint distribution $m_{ij}^{ST}$, we have:
\begin{equation}
\label{EdgeDensityCriterionEmpiricalFormula}
\begin{split}
c(M) = & -m \sum_{i = 1}^{k_S} {\sum_{j = 1}^{k_T} {\frac {m_{ij}^{ST}} {m} 
 \log \frac {\frac {m_{ij}^{ST}} {m}}   {\frac {m_{i.}^S} {m}  \frac {m_{.j}^T} {m}}   }} \\
&  - m \sum_{i = 1}^{n_S} {\frac {m_{i.}} {m} \log \frac {m_{i.}} {m}}
   - m \sum_{j = 1}^{n_T} {\frac {m_{.j}} {m} \log \frac {m_{.j}} {m}} \\
& + \mathcal{O}(\log m).  
\end{split}
\end{equation}

Considering that the empirical estimation asymptotically converges towards the related probabilities, the claim follows.
\end{proof}
\medskip

\noindent
{\bf Theorem~\ref{EdgeDensityCriterionLimit}}
{\it
The MODL approach for selecting a graph coclustering model $M$ asymptotically converges towards the true edge distribution, and the criterion for the best model $M_{\mbox{\footnotesize Best}}$ converges to $m$ times the entropy of the edge variable, which is the joint entropy of the source and target vertices variables.
\begin{equation*}
\lim_{m \rightarrow \infty} \frac {c(M_{\mbox{\footnotesize Best}})} {m} = H(V_S, V_T).
\end{equation*}
}

\begin{proof}
Using Theorem~\ref{EdgeDensityCriterionAsymptotics}, we have
\begin{equation*}
c(M) = - m I(V_S^M; V_T^M) + m H(V_S) + m H(V_T) + \mathcal{O}(\log m).
\end{equation*}

We apply the Data Processing Inequality (DPI) \cite{CoverEtAl91}, which states that post-processing cannot increase information.
More precisely, the DPI applies for three random variables $X, Y, Z$ that form a Markov chain $X \rightarrow Y \rightarrow Z$. It means that the conditional distribution of $Z$ depends only on $Y$ and is conditionally independent of $X$.
More specifically, for three random variables such that $p(Z | X, Y) = P(Z|Y)$, the DPI states that $I(X;Y) \geq I(X;Z)$

We apply the DPI to the variables $V_S, V_T, V_T^M$. As the vertex cluster variable $V_T^M$ can be computed according to a partition of the vertex variable $V_T$ ($V_T^M = f(V_T)$), we have $p(V_T^M|V_S, V_T) = p(V_T^M|V_T)$ and thus obtain:
\begin{equation}
I(V_S;V_T) \geq I(V_S;V_T^M).
\end{equation}

We apply again the DPI to the variables $V_T^M, V_S, V_S^M$. As the vertex cluster variable $V_S^M$ is a function of $V_S$, we have $p(V_S^M|V_S, V_T^M) = p(V_S^M|V_S)$ and get:
\begin{equation}
I(V_T^M;V_S) \geq I(V_T^M;V_S^M).
\end{equation}

By transitivity and since the mutual information is symmetrical, we get:
\begin{equation}
I(V_S;V_T) \geq I(V_S^M;V_T^M).
\end{equation}

It is noteworthy that this result applies to compare any pair of coclustering models, one of the models being a sub-partition of the other: the finer model brings a higher mutual information.

The model selection approach corresponds to a minimization of the MODL criterion. Let us now show that the best selected model asymptotically tends to be finer and finer, until reaching the finest possible model with one cluster per vertex, which is the maximal model $M_{\mbox{\footnotesize Max}}$ that enables a direct estimation of the edge probabilities $p_{ij}$:
\begin{equation}
I(V_S;V_T) = I(V_S^{M_{\mbox{\footnotesize Max}}};V_T^{M_{\mbox{\footnotesize Max}}}) \geq I(V_S^M;V_T^M).
\end{equation}

If $\forall \: M, I(V_S^{M_{\mbox{\footnotesize Max}}};V_T^{M_{\mbox{\footnotesize Max}}}) = I(V_S^M;V_T^M)$, then using Theorem~\ref{EdgeDensityCriterionAsymptotics}, the MODL approach asymptotically converges towards the true edge distribution.

\smallskip
If $\exists \: M_f, M_c, I(V_S^{M_{\mbox{\footnotesize Max}}};V_T^{M_{\mbox{\footnotesize Max}}}) = I(V_S^{M_f};V_T^{M_f}) > I(V_S^{M_c};V_T^{M_c})$, with $M_f$ a fine-grained model having the same mutual information as the maximal model and $M_c$ a coarse-grained model, then let us show that the approach asymptotically selects the fine-grained model $M_f$ rather than the coarser model $M_c$.

\smallskip
Let $\epsilon = \frac {I(V_S^{M_f};V_T^{M_f}) - I(V_S^{M_c};V_T^{M_c})} {2}$.

Using Theorem~\ref{EdgeDensityCriterionAsymptotics} for the convergence of the criterion for model $M_c$,
\begin{equation*}
\begin{split}
& \exists \: m_1 , \forall \: m \geq m_1,\\
& \quad \left| \frac {c(M_c)} {m} - \left( H(V_S) + H(V_T) - I(V_S^{M_c}; V_T^{M_c}) \right) \right| < \frac {\epsilon} {2}.
\end{split}
\end{equation*}

Similarly, for model $M_f$,
\begin{equation*}
\begin{split}
& \exists  \: m_2 , \forall \: m \geq m_2, \\
& \quad \left| \frac {c(M_f)} {m} - \left( H(V_S) + H(V_T) - I(V_S^{M_f}; V_T^{M_f}) \right) \right| < \frac {\epsilon} {2}.
\end{split}
\end{equation*}
 
Thus,
\begin{equation*}
\begin{split}
&\forall \: m \geq \max (m_1, m_2), \\
& \quad \frac {c(M_c)} {m} > H(V_S) + H(V_T) - I(V_S^{M_c}; V_T^{M_c}) - \frac{\epsilon} {2},\\
& \quad  \frac {c(M_f)} {m} < H(V_S) + H(V_T) - I(V_S^{M_f}; V_T^{M_f}) + \frac{\epsilon} {2}.
\end{split}
\end{equation*}

\begin{equation*}
\begin{split}
& \forall \: m \geq \max (m_1, m_2), \\
& \quad  \frac {c(M_f)} {m} - \frac {c(M_c)} {m} < I(V_S^{M_c}; V_T^{M_c}) - I(V_S^{M_f}; V_T^{M_f}) + \epsilon, \\
& \quad  \frac {c(M_f)} {m} < \frac {c(M_c)} {m} - \epsilon.
\end{split}
\end{equation*}

Since the model selection approach corresponds to a minimization of the MODL criterion, this means that the best selected model $M_{\mbox{\footnotesize Best}}$ asymptotically tends to be a fine-grained model $M_f$ having the same mutual information as the maximal model $M_{\mbox{\footnotesize Max}}$, which allows the estimation of the true edge distribution.
Using Theorem~\ref{EdgeDensityCriterionAsymptotics} with the maximum model (limit of the best selected model), we have:
\begin{equation*}
c(M_{\mbox{\footnotesize Max}}) = - m I(V_S; V_T) + m H(V_S) + m H(V_T) + \mathcal{O}(\log m).
\end{equation*}

As $I(X;Y) = H(X)+H(Y)-H(X,Y)$, the claim follows.
\end{proof}




%
\bibliographystyle{IEEEtran}
\bibliography{References}

\end{document}